\documentclass[12pt]{iopart}
\expandafter\let\csname equation*\endcsname\relax
\expandafter\let\csname endequation*\endcsname\relax
\usepackage[cmex10]{amsmath}
\usepackage[font=footnotesize]{subfig}
\usepackage{amssymb}
\usepackage[dvips]{graphicx}
\usepackage{fixltx2e}
\usepackage[abbr]{harvard}
\usepackage[export]{adjustbox}
\usepackage{textcomp}

\DeclareMathOperator{\argmin}{\arg\!\min}
\DeclareMathOperator{\DSC}{DSC}

\begin{document}

\title{Patch-based field-of-view matching in multi-modal images for electroporation-based ablations}
\author{Luc Lafitte$^{1}$, Rémi Giraud$^{2}$, Cornel Zachiu$^{3}$, Mario Ries$^{4}$, Olivier Sutter$^{5,6}$, Antoine Petit$^{5,6}$, Olivier Seror$^{5,6}$, Clair Poignard$^{1}$, Baudouin Denis de Senneville$^{1,3}$}

\address{$^1$ University of Bordeaux, IMB, UMR CNRS 5251, INRIA Project team Monc, Talence, France, F-33405 Talence Cedex, France}
\address{$^2$ University of Bordeaux, IMS, CNRS UMR 5218, F-33405 Talence Cedex, France}
\address{$^3$ Department of Radiotherapy, UMC Utrecht, Heidelberglaan 100, 3584 CX, Utrecht, The Netherlands}
\address{$^4$ Imaging Division, UMC Utrecht, Heidelberglaan 100, 3584 CX, Utrecht, The Netherlands}
\address{$^5$ Interventional radiology unit, H\^opitaux Universitaires Paris Seine Saint Denis, H\^opital Avicenne, Assistance Publique H\^opitaux de Paris, Bobigny France}
\address{$^6$ University of Paris 13, ``Sciences M\'edicale et Biologie Humaine'', Bobigny, France}

\begin{abstract}
Various multi-modal imaging sensors are currently involved at different steps of an interventional therapeutic work-flow. Cone beam computed tomography (CBCT), computed tomography (CT) or Magnetic Resonance (MR) images thereby provides complementary functional and/or structural information of the targeted region and organs at risk. Merging this information relies on a correct spatial alignment of the observed anatomy between the acquired images. This can be achieved by the means of multi-modal deformable image registration (DIR), demonstrated to be capable of estimating dense and elastic deformations between images acquired by multiple imaging devices. However, due to the typically different field-of-view (FOV) sampled across the various imaging modalities, such algorithms may severely fail in finding a satisfactory solution.

In the current study we propose a new fast method to align the FOV in multi-modal 3D medical images. To this end, a patch-based approach is introduced and combined with a state-of-the-art multi-modal image similarity metric in order to cope with multi-modal medical images. The occurrence of estimated patch shifts is computed for each spatial direction and the shift value with maximum occurrence is selected and used to adjust the image field-of-view. The performance of the proposed method --- in terms of both registration accuracy and computational needs --- is analyzed in the practical case of on-line irreversible electroporation procedures. In total, 30 pairs of pre-/per- operative IRE images are considered to illustrate the efficiency of our algorithm.

We show that a regional registration approach using voxel patches provides a good structural compromise between the voxel-wise and ``global shifts'' approaches. 
The method was thereby beneficial for CT to CBCT and MRI to CBCT registration tasks, especially when highly different image FOVs are involved. Besides, the benefit of the method for CT to CBCT and MRI to CBCT image registration is analyzed, including the impact of artifacts generated by percutaneous needle insertions. Additionally, the computational needs using commodity hardware are demonstrated to be compatible with clinical constraints in the practical case of on-line procedures. The proposed patch-based workflow thus represents an attractive asset for DIR at different stages of an interventional procedure.
\end{abstract}

\vspace{2pc}

\maketitle

\section{Introduction}

Multiple imaging devices can be involved at different stages of an interventional procedure, such as image-guided radiotherapy (IGRT) \cite{Guckenberger_2012}, irreversible electroporation (IRE) \cite{Gallinato_2019} or hyperthermia ablation \cite{MRgHIFU2} \cite{feedback}. In particular, cone-beam computed tomography (CBCT), computed tomography (CT) or Magnetic Resonance (MR) images are recently being employed at nearly all stages of the therapy: \emph{i.e.}, pre-, intra- and post-operatively. One of the benefits that employing multiple imaging sensors provides, is the ability to extract complementary functional and/or structural information of the targeted region and organs-at-risk. For example, as shown in \cite{Margin2016}, novel diagnostic indicators can thereby be calculated by fusing pre- and post-operative image data.
Similarly, multi-modal imaging may also be beneficial during the interventional procedure itself \cite{Zachiu2017a}. Of note is that the quality and the amount of data that can be acquired intra-operatively is generally limited by practical clinical considerations: CBCT guidance is for example particularly beneficial due to the low amount of imaging-related radiation delivered to the patient compared to a conventional CT scan. However, this often leads to the intra-operative images being subject to low contrast, low signal-to-noise ratio and artifacts. Thus, it would be of clinical benefit if such images would be augmented by pre-operative data \cite{Gallinato_2019}.
A common pre-requisite is that organ locations must be set in a common frame of reference. To this end, previous studies propose several multi-modal deformable image registration (DIR) algorithms dedicated to the estimation of dense and elastic deformations between images \cite{MIND,Rivaz_2014,EVolution2016}. This remains a challenging task since such algorithms have to be fast (to meet clinically acceptable durations) and automatic (the use must not be limited to a case-by-case basis and a manual recalibration is not preferable), especially when the patient is on the interventional table \cite{delineation} \cite{Zachiu2017b}.

A particular challenge arises when highly different fields-of-view (FOV) are sampled within the images. For example, while the FOV within intra-operative CBCT images is typically restricted to the targeted organ and its immediate surroundings, the corresponding pre-operative high-resolution CT image generally covers the entire abdomen and part of the thorax. A similar situation arises when a patient is screened via both CT/CBCT and MR imaging, with the resulting acquisitions typically having considerably different FOVs. This can severely hamper the performance of image registration algorithms, especially when iterative optimization strategies are employed: the algorithm is likely to get trapped into local optima if the apparent location of the anatomy-of-interest is too far apart within the two images. In such a case, a direct employment of DIR methods may be hardly feasible and a preliminary matching of the image FOVs (\emph{i.e.}, compensation of the 3D global shift between images) is necessary. 

While registration solutions optimizing a translational model may perform well for estimating rigid displacements, they may also become sub-optimal when elastic deformations are present between the images. Moreover, such methods typically imply high computational demands and manual tuning of several input parameters, which limits their use in a clinical setting \cite{Elastix2010}.

Alternatively, a regional registration approach using pixel/voxel patches, may provide a good structural compromise between the voxel-wise and ``global shifts'' approaches. Several patch or block-matching algorithms have been previously proposed, dedicated to various applications \cite{Jakubowski2013}. The aim of these approaches is to consider each pixel by its square neighborhood, to characterize its local context. Matching algorithms may then be used to find local correspondences between images.
 Nevertheless, these methods  are highly time consuming, especially when dealing with an important number of patches and when the search for correspondences must be performed in a large window search (\emph{i.e.}, searching for large patch displacements).
 A significant breakthrough has been obtained with the so-called ``PatchMatch'' algorithm \cite{PatchMatch}, which was initially proposed for finding pixel patch correspondences between 2D images in digital photography. The idea behind this approach is that some good patch matches can be found by random sampling, which can subsequently be allocated to surrounding areas as well, relying on the assumption that neighboring areas typically have similar displacements. The fast convergence of the process enables to quickly find good matches, even when these are located far from each other in the image spaces.
 This approach has been successfully employed to achieve numerous image analysis and editing tasks such as: 
 stereo matching \cite{bleyer2011patchmatch}, optical flow computation \cite{bao2014fast}, region inpainting \cite{newson2014video}, or 3D medical image segmentation \cite{giraud2016_nimg}.
 
In the current study, our contribution is four-fold:

\begin{enumerate}
  
 \item A new fast method --- using as a starting-point a 3D modified PatchMatch algorithm --- is proposed to align the FOV in medical 3D images. A user-defined mask surrounding a region/organ of interest in one of the images can be provided as an input so that a global shift can be estimated relying on image information from this specific region.
 
 \item The modified PatchMatch algorithm is  combined with a well-adapted multi-modal image similarity metric in order to cope with multi-modal medical images.
 
 \item The performance --- in terms of registration accuracy and computational needs --- is analyzed, and demonstrated to be compatible with clinical constraints in the practical case of on-line irreversible electroporation procedures. In total, 30 pairs of pre-/per- operative IRE images are considered to illustrate the efficiency of our algorithm.
 
 \item The benefit of proposed approach is evaluated for a potential pre-conditioning of a more complex multi-modal DIR algorithm.
 
\end{enumerate}

\section{Materials and Methods}

\subsection{Proposed method}

Let $I$ and $J$ be two 3D images. In the scope of this study, let $I$ and $J$ be a pre- and an intra-operative image, respectively. We seek X-, Y- and Z- translation components between $I$ and $J$ in order to match the position an organ of interest manually delineated in $I$. 
We recall that the estimation of elastic organ deformations by itself is outside the scope of the study: the proposed workflow is solely intend to standardize field-of-view in multi-modal images for a potential pre-conditioning a more complex multi-modal elastic registration algorithm. 

\begin{figure}[h!]
\begin{minipage}[b]{\linewidth}
\centering
\centerline{\includegraphics[trim={0cm 0cm 0cm 0cm},clip,width=\linewidth]{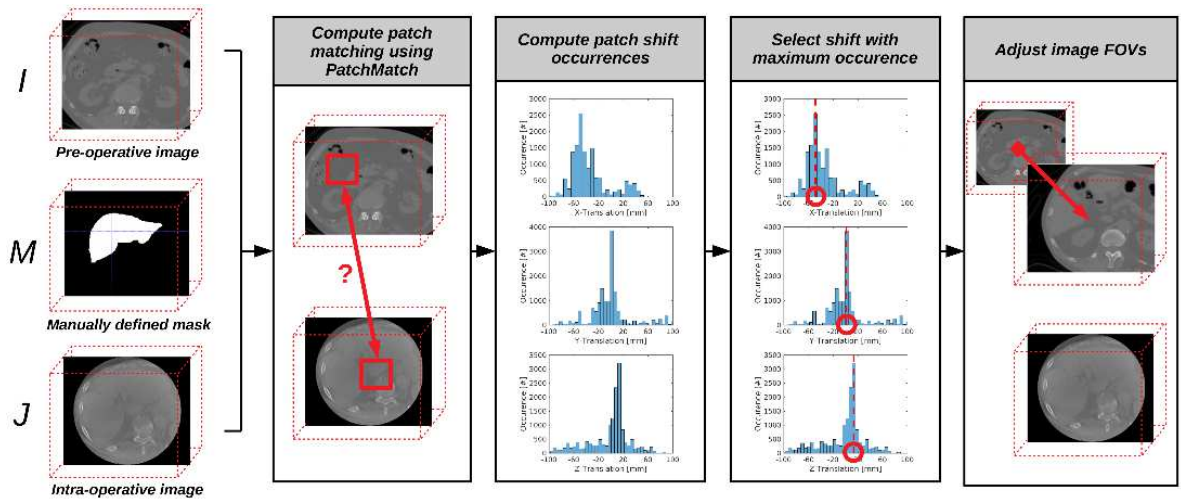}}
\end{minipage}
\caption{Data processing sequence designed for the fast standardization of field-of-view in multi-modal images using the proposed patch-based framework.}
\label{fig:scheme}
\end{figure}

The proposed method (detailed in Figure \ref{fig:scheme}) includes the following three main successive steps:

\begin{enumerate}
 \item The PatchMatch (PM) algorithm is adapted and combined with a multi-modal metric in order to compute patch correspondences between $I$ and $J$ (see section \ref{sssec::PM}). The multi-modal metric aims at evaluating edge alignments (EA) within patches.
 \item The occurrence of estimated patch shifts, 
 \emph{i.e.}, the displacement between the patch positions in $I$ and their correspondence in $J$, is computed for each spatial direction (see section \ref{sssec::metric}).
 \item For each spatial direction, the shift value with maximum occurrence is selected and used to adjust the image field-of-view (see section \ref{sssec:histo}).
\end{enumerate}

The proposed method is referred to as ``PM-EA'' (PatchMatch-Edge Alignment) in the scope of this study.

\subsubsection{Manual delineation of the organ of interest.}
\label{sssec::masking}

The pre-operative image $I$ is first used to manually segment the targeted region of interest. A binary mask (denoted by $M$) is constructed. Voxels of the image inside the mask have a value of one, and outside a value of zero. We underline that, in the scope of this study, this process was done using the pre-operative image $I$ only, in order to demonstrate that the method is compatible with an automatic use during an intra-operative session. Note that this delineation step is often performed anyway during the planning session of the therapy and thus does not put extra burden on the medical staff.

\subsubsection{Preprocessing of input data.}

$I$, $J$ and $M$ were resampled onto a common grid with a voxel size of $1 \times 1 \times 1$ millimeters using a trilinear interpolation. 

\subsubsection{Implemented PatchMatch algorithm.}
\label{sssec::PM}

PatchMatch is an iterative algorithm designed to quickly estimate patch correspondences between two given 2D images \cite{PatchMatch}. 
In our study, we first extend this algorithm for the matching of 3D patches. 
Hence, a patch consists in a cubic subset of the image domain, denoted by $\Gamma$, centered on one single voxel. Let $\vec{r}=(x,y,z)\in \Omega$ be the spatial location of the center voxel, $\Omega$ the image domain and $(x,y,z)$ the voxel coordinates.

\begin{itemize}
\item Initialization. An initial guess is first computed: each patch from image $I$ is initially randomly matched with a patch from image $J$. Subsequently, at each iteration of PatchMatch, voxels are scanned from left to right (X-axis), head to foot (Y-axis), front to back (Z-axis). For each voxel examination, the corresponding patch undergoes a \textit{``propagation step''} followed by a \textit{``random search step''}, as described in the seminal paper \cite{PatchMatch}:
The output of our algorithm is a patch shift map $\vec{V}$, defined for each voxel in $\Omega$.

\item{Propagation step.}
During this step, in order to find better correspondences, the patch shift of the current voxel $\vec{r}$ in $I$ in $I$ is considered to be similar to the ones of its three already examined neighboors in each direction (6-connexity) (\emph{i.e.}, the three voxels at locations $\vec{r}_{(-1,0,0)}=(x-1,y,z)$, $\vec{r}_{(0,-1,0)}=(x,y-1,z)$ and $\vec{r}_{(0,0,-1)}=(x,y,z-1)$). 

Let $\vec{V}=(u,v,w)$ be the patch shifts that we seek ($(u,v,w)$ being the voxelwise patch shift coordinates). Let $\vec{r}_1$ and $\vec{r}_2$ be two given spatial location in $\Omega$ and $D(\vec{r}_1,\vec{V}(\vec{r}_2))$ the distance between the patch at location $\vec{r}_1$ in $I$ and the patch at location $\vec{r_1}+\vec{V}(\vec{r}_2)$ in $J$. $\vec{V}(\vec{r})$ is updated as follows: 

\begin{equation}
 \vec{V}(\vec{r})=\argmin _{\vec{V}} \{ D(\vec{r},\vec{V}(\vec{r})), D(\vec{r},\vec{V}(\vec{r}_{(-1,0,0)})), D(\vec{r},\vec{V}(\vec{r}_{(0,-1,0)})), D(\vec{r},\vec{V}(\vec{r}_{(0,0,-1)})) \}
 \label{eq:patch_correspondence}
\end{equation}

\item{Random search step.} 
This step attempts to improve $\vec{V}(\vec{r})$ by computing a set of candidate shifts (noted $\vec{V}_i(\vec{r})$) at an exponentially decreasing spatial distance from $\vec{V}(\vec{r})$:

\begin{equation}
\vec{V}_i(\vec{r}) = \vec{V}(\vec{r}) + w \alpha ^i R_i 
\end{equation}

\noindent $R_i$ being a uniform random in $\left[-1, 1\right]^3$, $w$ the maximum search distance (set to the maximum image dimension), and $\alpha$ a fixed ratio between search window sizes (we took $\alpha=0.5$, as suggested in \cite{PatchMatch}). Patches for $i=$ 0, 1, 2 and so on  are examined until the current search distance $w \alpha^i$ falls below one voxel.
\\

As reported in Barnes et al seminal paper, PatchMatch provides satisfactory results using a fixed number of iterations (5 iterations max) and that the algorithm converges most rapidly in the first iterations \cite{PatchMatch}. In the current study we used two iterations, since it was found to be a good compromise between accuracy and computational costs.

In our implementation, the input images are down-sampled before PatchMatch: while lower computation times are expected using down-sampled versions of input images, it should also impact overall registration results. The down-sampling factor is thus an important input parameter for the algorithm and its impact will be carrefully analysed and discussed below.

To further reduce the computational burden, the search window were a rectangular bounding box including both $J$ and voxels with a value of one in $M$.

\item{Aggregation of multiple PM estimations.}
As for exemplar-based segmentation \cite{giraud2016_nimg}, our method can benefit from multiple PM estimations. Patch-shift estimates indeed rely on random candidate selection and several independent processes may provide different correspondences. Although PatchMatch inherently relies on serial operations and cannot benefit from parallel architectures in its current form, multiple realisations of PatchMatch can easily be calculated using separate CPU threads. Let $V^j(\vec{r})=(u^j(\vec{r}),v^j(\vec{r}),w^j(\vec{r}))$ be one realisation at spatial location $\vec{r}$, $j$ being the realisation index. 
For each voxel, the obtained 3D shift realisations were then combined into a single 3D median vector $V(\vec{r})$ \cite{VN1990} as follows:

\begin{equation}
V(\vec{r}) =  \argmin_{V(\vec{r}) \in \{ V^j(\vec{r}) \}} \left( \sum_j \left| V(\vec{r})-V^j(\vec{r}) \right| \right)\\
\end{equation}

In this manner, outliers realisations were discarded without any additional penalty on the computational burden.
\end{itemize}
\subsubsection{Proposed multi-modal metric.}
\label{sssec::metric}

In the original PatchMatch paper, the distance between patches $D(.)$ is the Sum of Squared Differences (SSD, also referred to as L2-norm) applied on the image intensity. While for mono-modal registration algorithms the SSD applied directly on the images might be sufficient \cite{HS1981} \cite{pca}, such a measure is unsuitable for registering across modalities. A modality independent similarity measure is thus necessary. 
In the current paper we used an existing multi-modal metric which favors edge alignments (EA) in both patches \cite{EA1998} \cite{Sutour2015}.

Let $\vec{\nabla} _I$ and $\vec{\nabla} _J$ be the gradient of the reference image $I$ and the image to register $J$, respectively. The distance $D(\vec{r}_1,\vec{V}(\vec{r}_2))$ between a patch of interest $\Gamma$ in $I$ (centered on the voxel located in $\vec{r}_1$) and its potential correspondence in $J$ (centered on the voxel located in $\vec{r}_1+\vec{V}(\vec{r}_2)$) was defined as follows:

\begin{equation}
D(\vec{r}_1,\vec{V}(\vec{r}_2)) = - \frac{\int \limits_{\Gamma} \left| \vec{\nabla} _I \left( \vec{r}_1 \right) \cdot \vec{\nabla} _J \left( \vec{r}_1 + \vec{V}(\vec{r}_2) \right) \right| \mathrm{d}\vec{r}}
{\int \limits_{\Gamma} \left\lVert \vec{\nabla} _I \left( \vec{r}_1 \right) \right\rVert_2 \left\lVert \vec{\nabla} _J \left( \vec{r}_1 + \vec{V}(\vec{r}_2) \right) \right\rVert_2 \mathrm{d}\vec{r}}
\label{eq:metric}
\end{equation}

\noindent where $\| \cdot \| _2$ is the Euclidean norm.

Practically, the scalar product in the numerator is maximized when the edges in $\Gamma$ are aligned with edges in the potential corresponding patch in $J$. Note that the numerator is maximized regardless any possible contrast reversals: due to the absolute value, the numerator is maximized for both parallel and anti-parallel edges. In addition, the scalar product in the numerator favors strong edges present in both modalities. The denominator, for its part, acts as a normalisation factor. Ultimately, since a minimization of $D$ is required to compute patch correspondences in Eq. (\ref{eq:patch_correspondence}), a negative sign has been set behing the fractional term in Eq. (\ref{eq:metric}).

\subsubsection{Matching image FOVs from estimated patch correspondences.}
\label{sssec:histo}

At this point we have a voxelwise 3D shift maps $\vec{V}(\vec{r})$, $\vec{r} \in \Omega$. The objective is now to simplifiy $\vec{V}$ down to a single 3D image shift. 
For this purpose, we individually analysed shift occurrences in each component of $\vec{V}$ (\emph{i.e.}, $u$, $v$ and $w$) and within the binary mask $M$ encompassing the target organ: for each component, a histogram was calculated and the shift value with the highest occurrence was selected. The obtained 3D image shift was subsequently used to adjust the FOV of $J$ with respect to the one of $I$.

Note that lower and upper bounds on estimated shift values as well as a number of bins needs to be determined for the construction of the histograms. The lower and upper bounds were set to -100 and 100 millimeters, respectively, which was found to be sufficient in our tests. The number of bins is an input parameter for the algorithm that will be analysed below.

\subsection{Experimental evaluation}

\subsubsection{Data sets.}
\label{sssec:data}

For the evaluation of the method we used data acquired during IRE procedures which are routinely performed at the University Hospital Jean Verdier at Bondy in France. This retrospective study is in accordance with ethical principals of the Declaration of Helsinki and has been approved by the local committee on human research of the University Hospital J Verdier. 

The clinical workflow included the following sessions:
\begin{itemize}
\item Pre-operative session. This session, performed several days before the interventional procedure, allowed identifying the tumor and the main liver structures using either a CT-scan (voxel size=$[0.67-0.88]\times[0.67-0.88]\times[1.25-2]$ mm$^3$, FOV=$[341-450]\times[341-450]\times[182-506]$ mm$^3$) or a MR-scan (T1-weighting, voxel size=$1.72\times1.72\times[2.5-3]$ mm$^3$, FOV=$440\times440\times[180-200]$ mm$^3$).
 
\item {Interventional session.} The day of the procedure, an IRE ablation was performed under general anesthesia. The needles are percutaneously inserted around the tumor by the interventional radiologist with a free-hand technique under combination of real-time ultrasound (US) and 3D Virtual Target Fluoroscopic Display such that the electric field covers the target region \cite{Fluoroscopic2018}. A 3D CBCT (voxel size=$0.45\times0.45\times0.45$ mm$^3$, FOV=$230\times230\times[192-256]$ mm$^3$) imaging was performed to visualise liver and needle locations. During the interventional session, a registration with pre-operative data is intend to augment the CBCT with tumor/liver structures segmentations. The pursued objective is to improve the targeting and to allow dose modeling, as described in \cite{Gallinato_2019}.
\end{itemize}

In total, we analysed a set of 30 pairs of pre-/intra-operative images distributed over the four following groups: 

\begin{enumerate}
 \item 8 pairs of CT/CBCT images obtained on 8 patients, respectively. CBCTs were acquired before needle insertion.
 \item 8 pairs of CT/CBCT images. Patients and CTs were those used in (i). CBCTs were acquired after needle insertion.
 \item 7 pairs of MR/CBCT images obtained on 8 patients, respectively. CBCTs were acquired before needle insertion.
 \item 7 pairs of MR/CBCT images. Patients and MRs were those used in (iii). CBCTs were acquired after needle insertion.
\end{enumerate}

\subsubsection{Performance assessment.}

The Dice Similarity Coefficient (DSC) was employed to determine the contour overlap of the liver:

\begin{equation}
\DSC = \frac{2\left|A \cap B \right|}{\left|A \right| + \left|B \right|}
\end{equation}

\noindent where $A$ and $B$ are two manually defined ROIs encompassing the liver in the reference and the corrected image, respectively. $A \cap B$ is their intersection and $\left| \cdot \right|$ denotes the cardinality of a set (\emph{i.e.}, the number of voxels).

$\DSC$ mean and standard deviation were computed over the 4 sets of image pairs individually (\emph{i.e.}, CT/CBCT no needle, CT/CBCT needles, MR/CBCT no needle, MR/CBCT needles, as defined in section \ref{sssec:data}). A Wilcoxon paired test was carried out in order to study whether $\DSC$ differences are statistically significant. A significance threshold of $p = 0.05$ was used.

\subsubsection{Calibration of the proposed PM-EA algorithm.}
\label{sssec:calib}

At this point it is important to underline that four main input parameters may influence the performance of the proposed approach. The proposed PM-EA algorithm was challenged against various modifications applied to these calibration parameters:

 \paragraph{Input parameter $\#1$: down-sampling of input images (section \ref{sssec::PM}).} $\DSC$ and computation time were calculated using $I$, $J$ and $M$ at original image dimension and using down-sampling (DS) factors $2\times$, $4\times$ and $8\times$ in each spatial direction. A default down-sampling factor of $8\times$ was used.
 
 \paragraph{Input parameter $\#2$: patch size (section \ref{sssec::metric}).} $\DSC$ and computation time were calculated using patches of size $3\times 3\times 3$, $5\times 5\times 5$, $7\times 7\times 7$, $9\times 9\times 9$ and $11\times 11\times 11$ voxels. A default size of $9\times 9\times 9$ voxels was used.
 
 \paragraph{Input parameter $\#3$: number of histogram bins (section \ref{sssec:histo}).} $\DSC$ were calculated for number of histogram bins of 10, 30, 50, 70 and 90. The computation time was not evaluated since a marginal impact is expected here. A default value of 50 was used.

 \paragraph{Input parameter $\#4$: manual delineation errors of the targeted organ (section \ref{sssec::masking}).} To analyse potential errors arising from the manual delineation process, $M$ was iteratively eroded (resp. dilated) using a $5 \times 5 \times 5$ kernel. At each iteration, the volume of the eroded (resp. dilated) mask was calculated as well as the corresponding DSC after image alignement. Here again, the computation time was not evaluated since a marginal impact is expected.

\subsubsection{Tested algorithms.}
\label{sssec:tested_algos}

The PM-EA's ability to estimate a global 3D translation between images was challenged against two competing approaches. We also evaluated the benefit of using PM-EA as a starting point for an existing more complex multi-modal elastic registration algorithm. The above-mentioned 30 pairs of images were processed using PM-EA and using the following selection of image registration solutions:

\paragraph{Elastix.} We have selected in the Open source Elastix registration software \cite{Elastix2010} \cite{Shamonin2013} a registration solution which employs a 3D translation transformation model and which maximizes the normalized mutual information \cite{Studholme1999} between the images to be registered. This registration solution is referred to as ``Elastix'' in the following.

\paragraph{PM-L2.}  The proposed PatchMatch registration framework is here employed using a L2-norm for patch comparison, as it was introduced in the original paper \cite{PatchMatch}. This registration solution is referred to as ``PM-L2'' hereafter.

\paragraph{Evo.} The EVolution algorithm (which is abbreviated as ``Evo'' in the scope of this study) was employed to estimate the elastic deformation $\vec{V}$ as the minimizer of the following energy $E$:
\begin{equation}
E(\vec{V}) = \int \limits_{\Omega} \exp(D(\vec{V})) + \frac{\alpha}{2} \left( \parallel\vec{\nabla} u \parallel _2 ^2 + \parallel \vec{\nabla} v \parallel _2 ^2 + \parallel \vec{\nabla} w \parallel _2 ^2 \right) \mathrm{d}\vec{r},
\label{eq:edge_functional}
\end{equation}
 $D(\vec{V})$ being the multi-modal metric of Eq. (\ref{eq:metric}) and $\alpha$ a weighting factor designed to link both the data fidelity term (left part of Eq. (\ref{eq:edge_functional})) and the motion field regularity (right part of Eq. (\ref{eq:edge_functional})). Note that $D(\vec{V})$ is composed with an exponential function in order to ensure that the data fidelity term is a positive-definite function. Additional details concerning the manner in which the Evo functional is minimized together with a detailed analysis of the algorithm performance can be found in \cite{EVolution2016}. 
A numerical implementation designed for the reduction of computational costs can be found in \cite{SV_Evo2018}.

\paragraph{PM-EA+Evo.} The above-mentioned multi-modal registration algorithm Evo is here employed after FOV standardization using proposed PM-EA algorithm. The registration workflow, comprised of PM-EA followed by Evo, is referred to as ``PM-EA+Evo'' throughout the rest of the manuscript.

Each of Elastix, PM-L2 and PM-EA aims to estimate a global 3D translation between $I$ and $J$ within the shortest possible time. A common factor of $8\times$ in each spatial direction (i.e. the default value for PM-EA given in section \ref{sssec:calib}) was used to down-sample the input data (\emph{i.e.}, $I$, $J$ and $M$). 

On the other hand, elastic registration is known to be a complex task. Using Evo, a down-sampling of input data (\emph{i.e.}, $I$, $J$ and $M$) by a factor $4\times$ was used in order to maintain registration accuracy of the outputs and computation times compatible with our clinical constraints (below 30 seconds).

\subsection{Hardware and implementation}

Our test platform was an Intel 2.5 GHz i7 workstation (8 cores) with 32 GB of RAM. 
The implementation was performed in C++ and parallelized through multi-threading (one thread per core).

\section{Results}

\begin{figure}[h!]
\begin{minipage}[b]{0.16\linewidth}
\centering
 \centerline{\small{Before needle}}\medskip
 \vspace{-0.2cm}
 \centerline{\small{insertion}}\medskip
 \vspace{2.2cm}
\end{minipage}
\begin{minipage}[b]{0.25\linewidth}
\centering
\centerline{\footnotesize{Transversal}}\medskip
\centerline{\includegraphics[trim={0cm 0cm 0cm 0cm},clip,height=3.8cm]{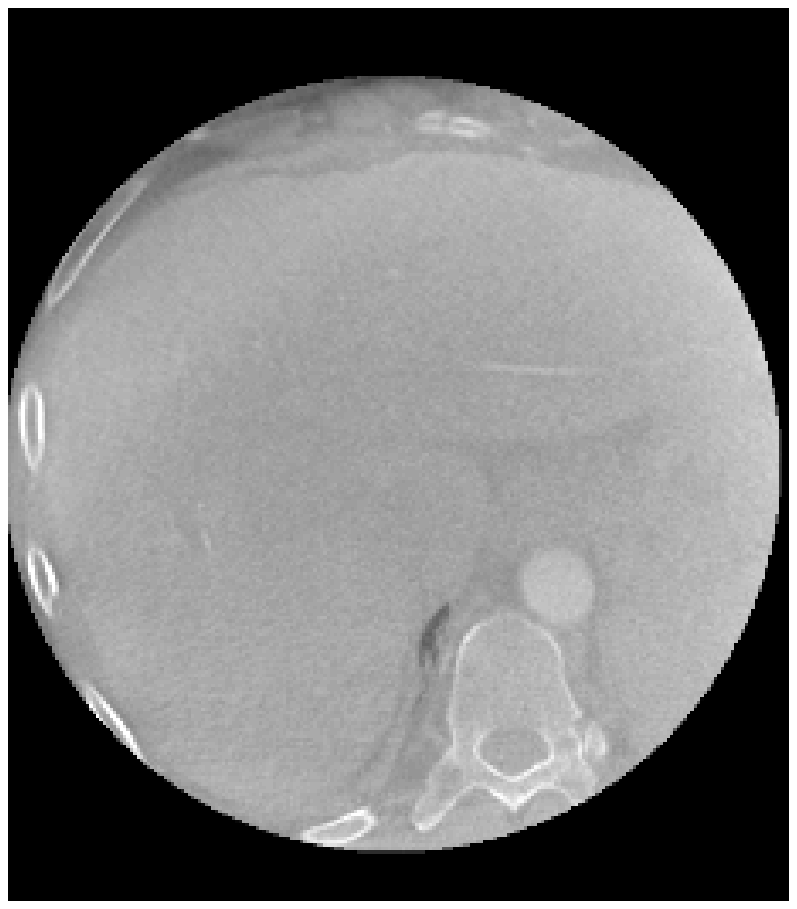}}
\centerline{(a)}\medskip
\end{minipage}
\begin{minipage}[b]{0.27\linewidth}
\centering
\centerline{\footnotesize{Sagittal}}\medskip
\centerline{\includegraphics[trim={0cm 0cm 0cm 0cm},clip,height=3.8cm]{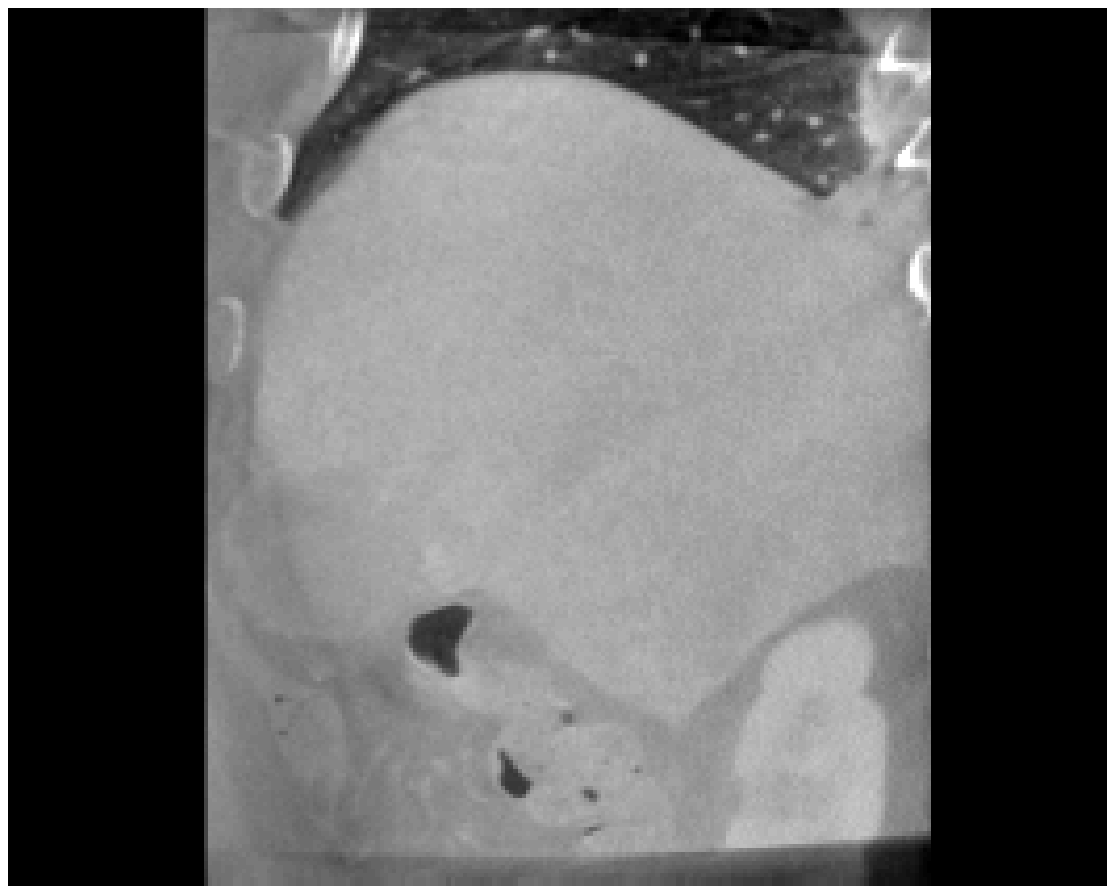}}
\centerline{(b)}\medskip
\end{minipage}
\begin{minipage}[b]{0.3\linewidth}
\centering
\centerline{\footnotesize{Coronal}}\medskip
\vspace{0.08cm}
\centerline{\includegraphics[trim={0cm 0cm 0cm 0cm},clip,height=3.8cm]{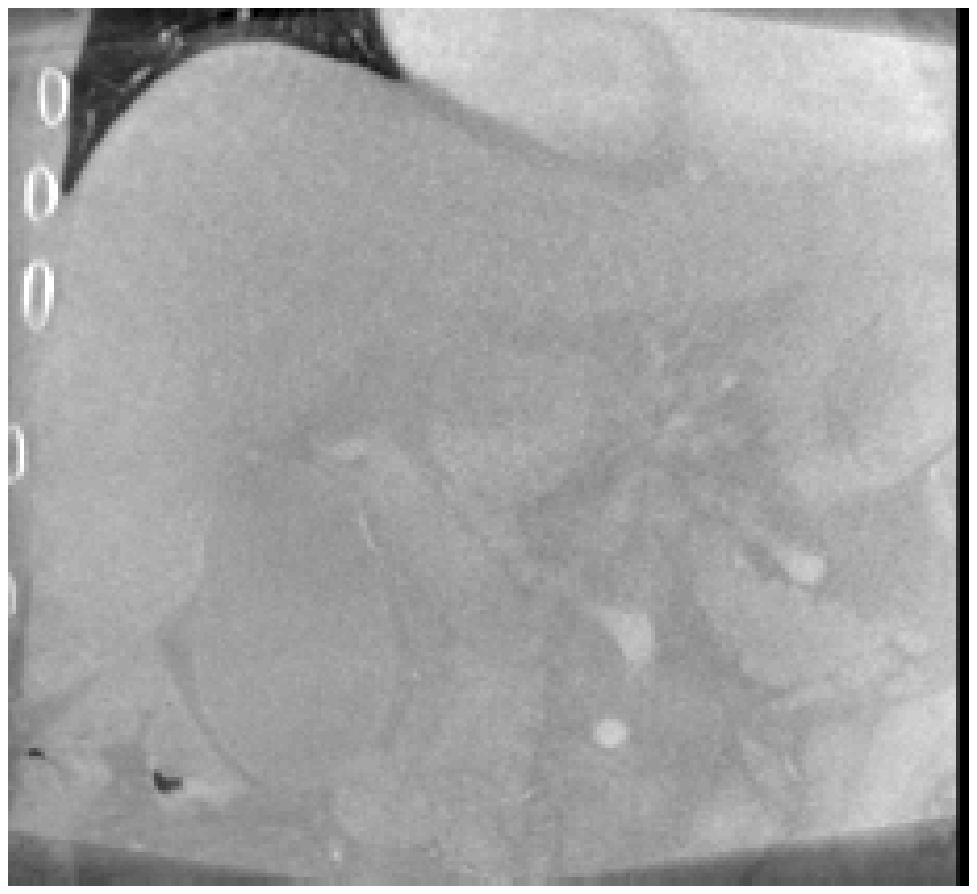}}
\centerline{(c)}\medskip
\end{minipage}

\begin{minipage}[b]{0.16\linewidth}
\centering
 \centerline{\small{After needle}}\medskip
 \vspace{-0.2cm}
 \centerline{\small{insertion}}\medskip
 \vspace{2.2cm}
\end{minipage}
\begin{minipage}[b]{0.25\linewidth}
\centering
\centerline{\includegraphics[trim={0cm 0cm 0cm 0cm},clip,height=3.8cm]{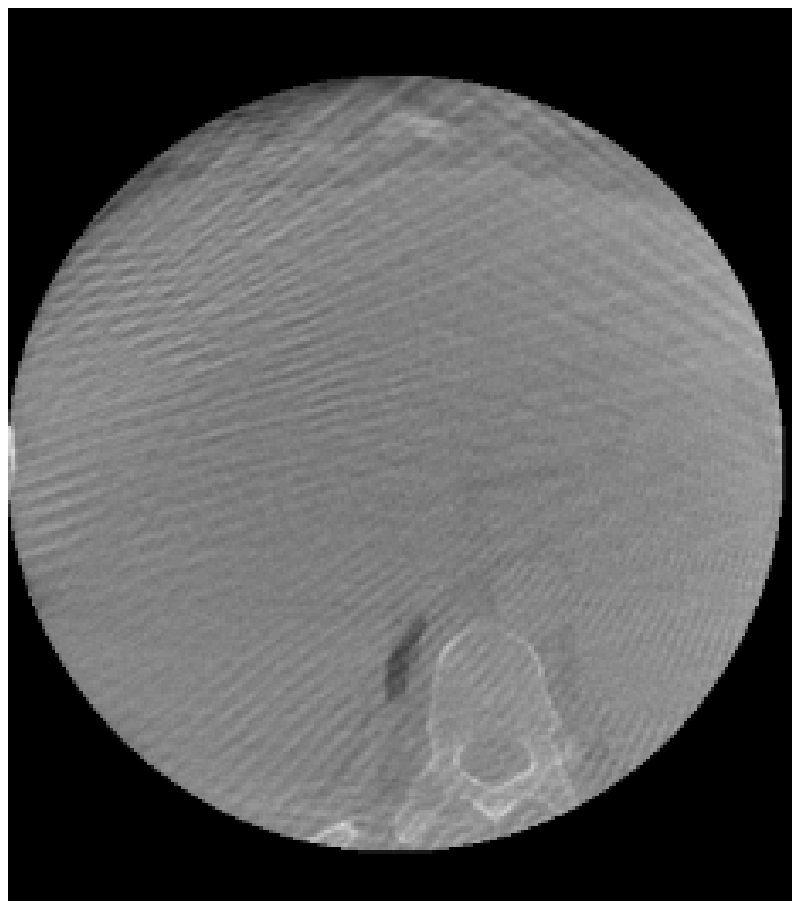}}
\centerline{(d)}\medskip
\end{minipage}
\begin{minipage}[b]{0.27\linewidth}
\centering
\centerline{\includegraphics[trim={0cm 0cm 0cm 0cm},clip,height=3.8cm]{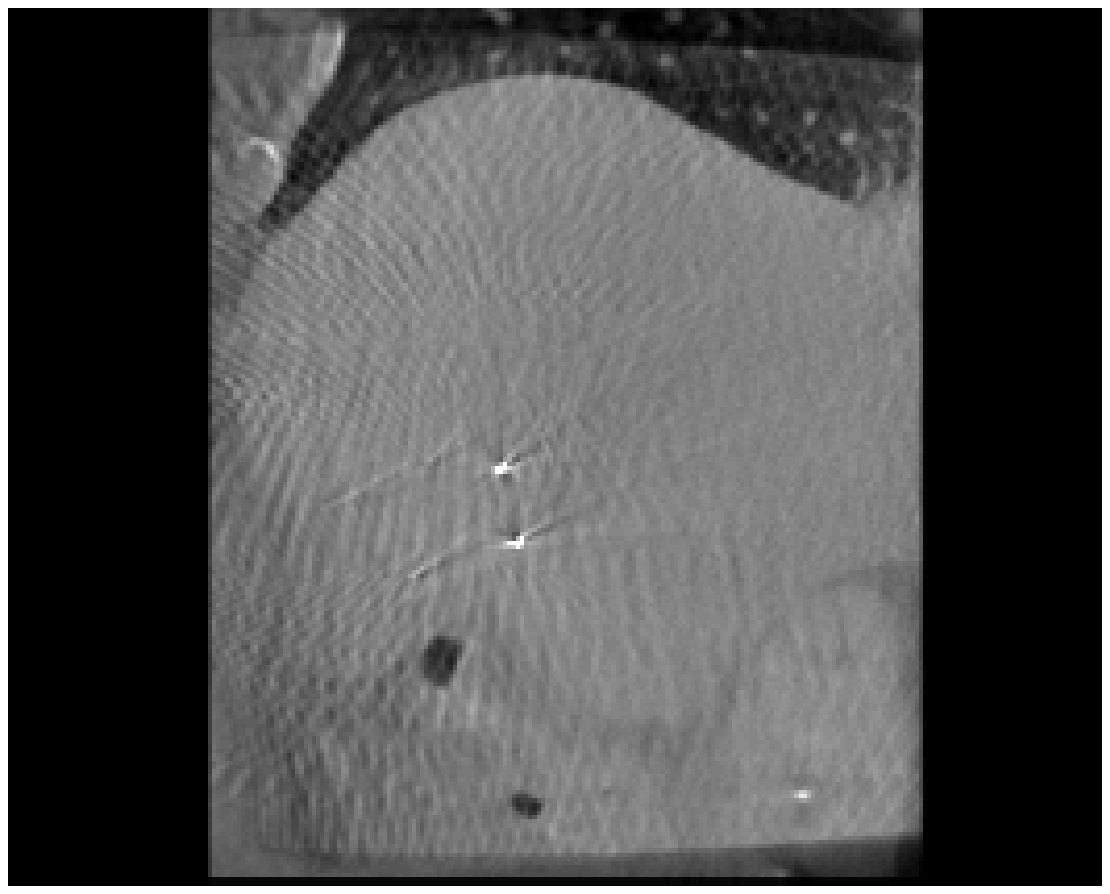}}
\centerline{(e)}\medskip
\end{minipage}
\begin{minipage}[b]{0.3\linewidth}
\centering
\vspace{0.08cm}
\centerline{\includegraphics[trim={0cm 0cm 0cm 0cm},clip,height=3.8cm]{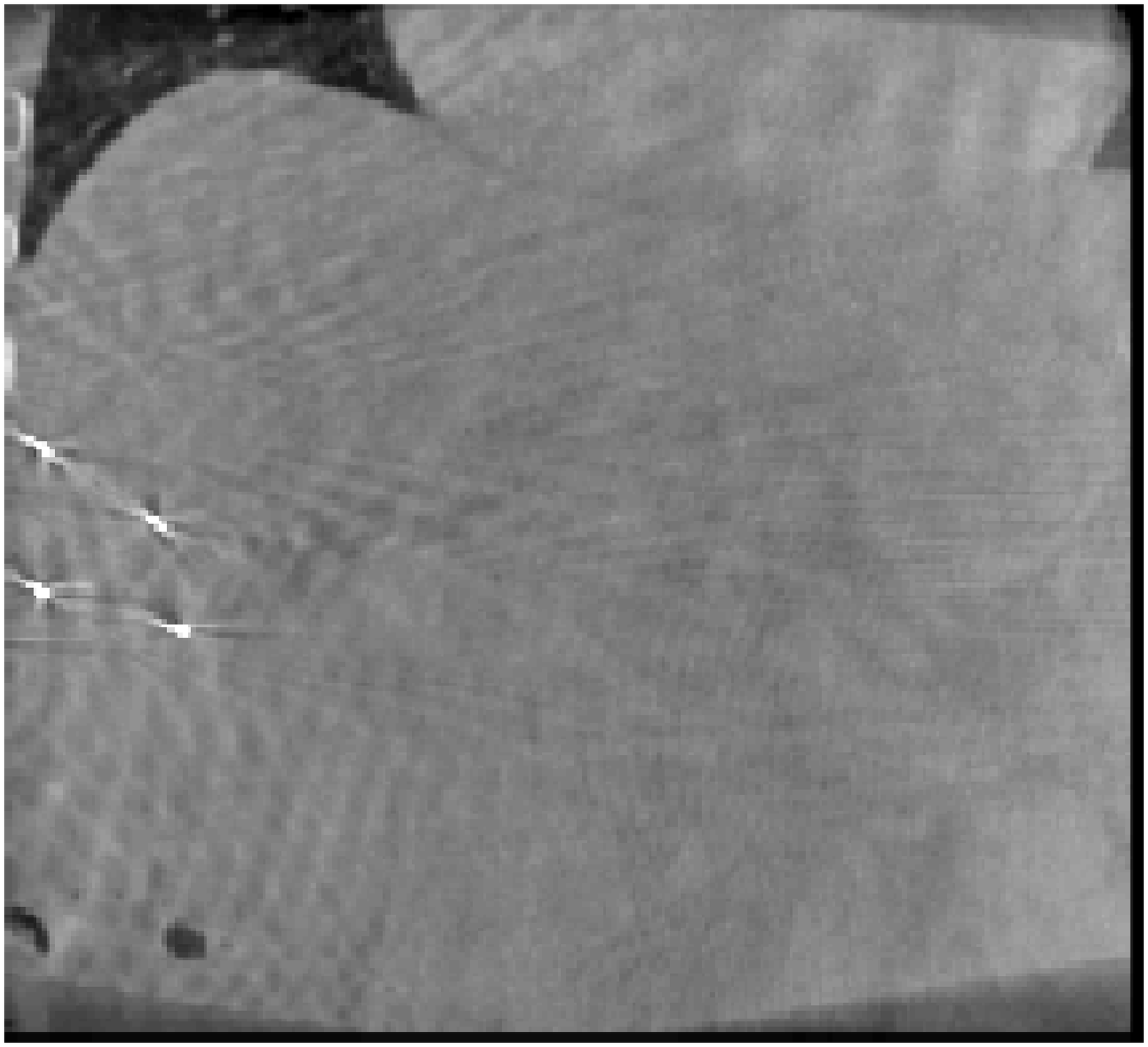}}
\centerline{(f)}\medskip
\end{minipage}
\caption{Typical CBCT images obtained during an IRE procedure. Compared to the image acquired before needle insertion (top row), the image acquired after the insertion of four needles (bottom row) is visibly altered by streaking artifacts. The latter introduce intensity variations which obstructs and degrades finer details of the anatomy. The partial image FOV is also observable: only data contained within a cylinder are available (see the circular FOV in the transveral plane in the first column).}
\label{fig:CBCT_images}
\end{figure}

Figure \ref{fig:CBCT_images} provides a visual assessment of challenges arising from the use of CBCT during the intra-operative IRE session. Middle transversal, sagittal and coronal slices are illustrated for CBCT images acquired on the same patient before and after insertion of 4 needles. First, only voxels contained within a cylinder have non-zero values: a partial circular FOV in the transveral plane is observable in the first column (see \ref{fig:CBCT_images}a and \ref{fig:CBCT_images}d). In addition, a low contrast-to-noise ratio is observable on both images. Also, an increasing amount of streaking artifacts is observable after needle insertions.

\begin{figure}[h!]
\begin{minipage}[b]{0.08\linewidth}
\centering
 \centerline{\small{Trans.}}\medskip
 \vspace{-0.2cm}
 \centerline{\small{[X-Y]}}\medskip
 \vspace{2.2cm}
\end{minipage}
\begin{minipage}[b]{0.22\linewidth}
\centering
\centerline{\footnotesize{CBCT}}\medskip
\vspace{0.08cm}
\centerline{\includegraphics[trim={0cm 0cm 0cm 0cm},clip,width=3.6cm]{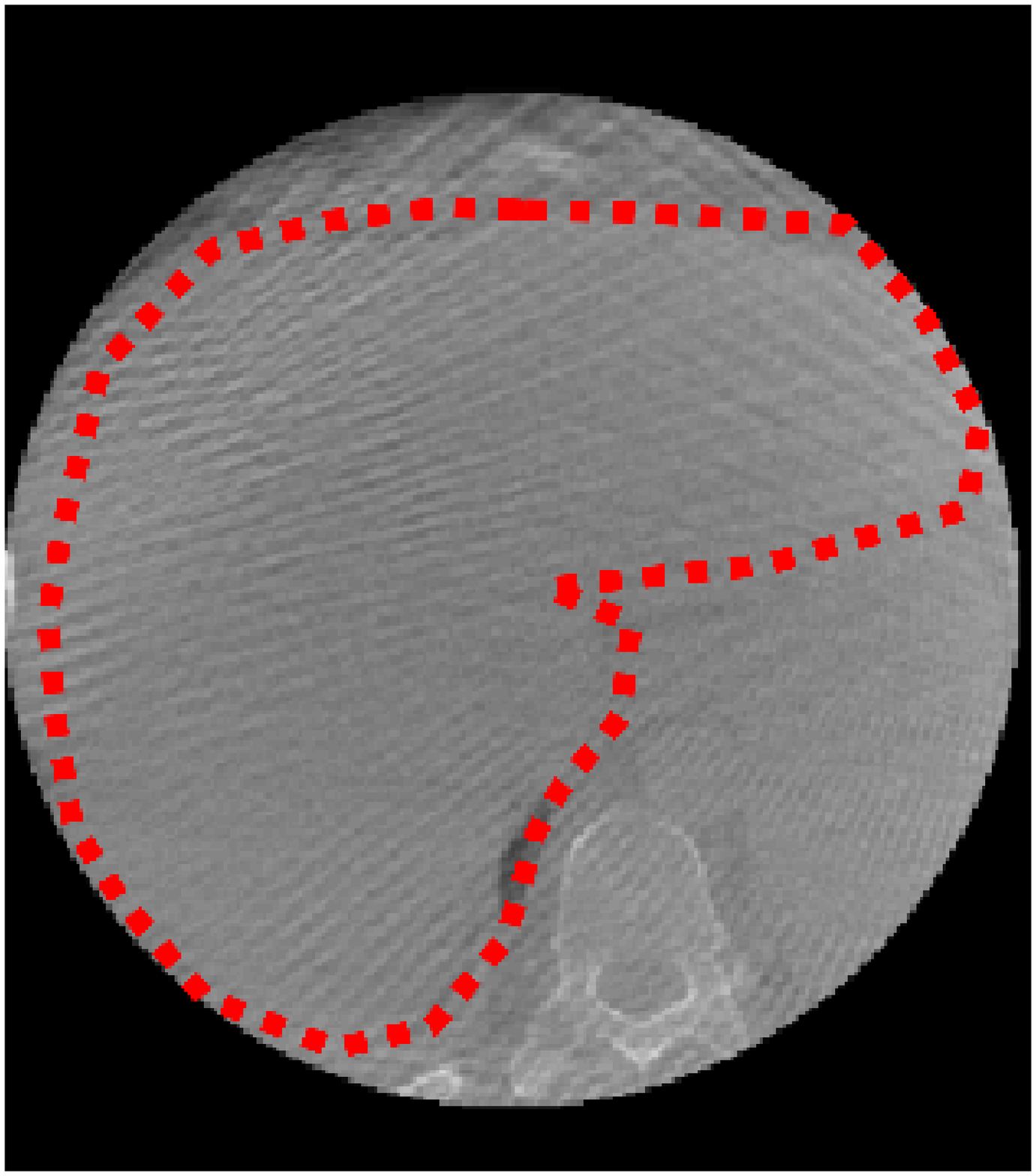}}
\centerline{(a)}\medskip
\end{minipage}
\begin{minipage}[b]{0.22\linewidth}
\centering
\centerline{\footnotesize{CT / No registration}}\medskip
\centerline{\includegraphics[trim={0cm 0cm 0cm 0cm},clip,width=3.6cm]{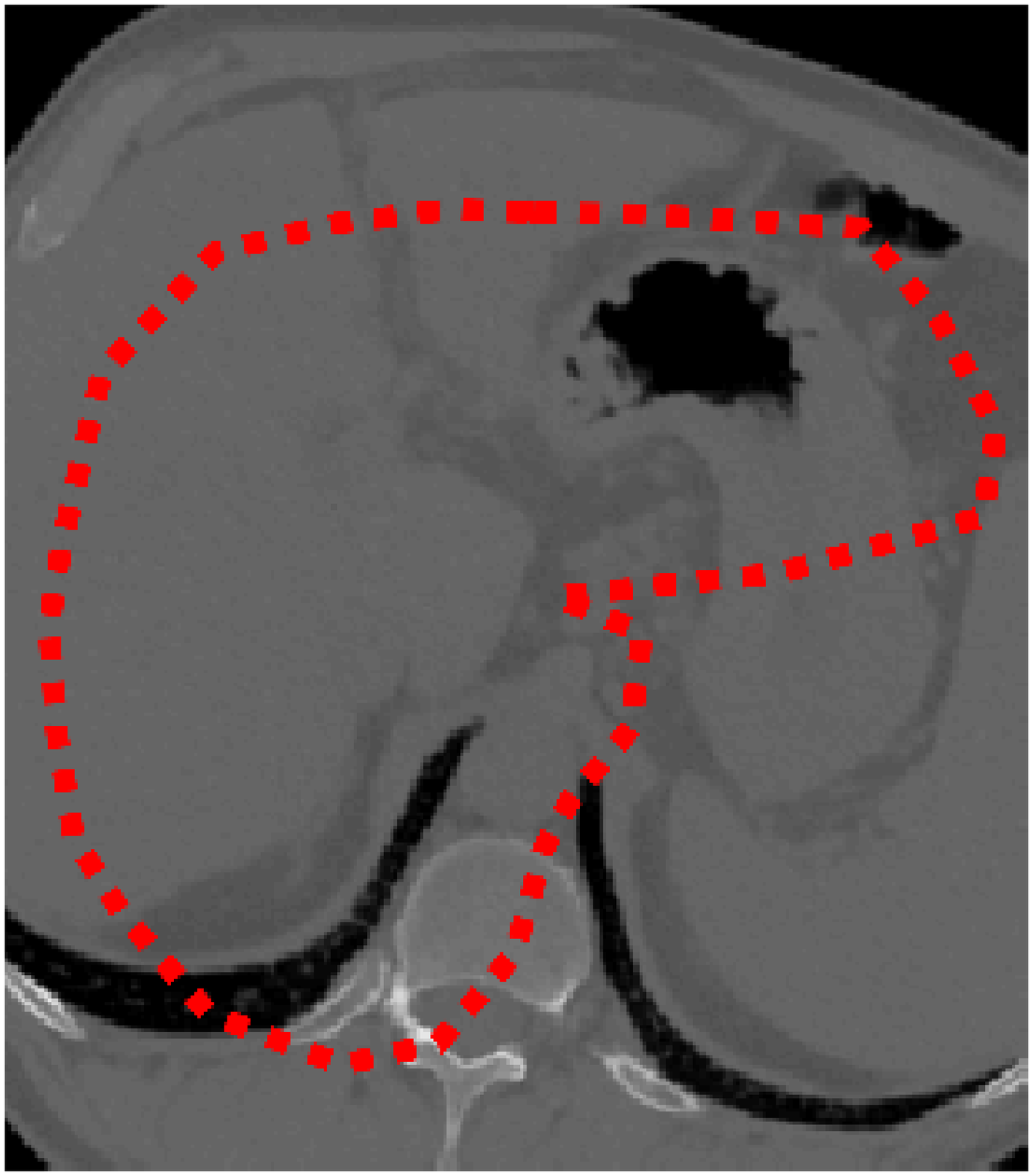}}
\centerline{(b)}\medskip
\end{minipage}
\begin{minipage}[b]{0.22\linewidth}
\centering
\centerline{\footnotesize{CT / PM-EA}}\medskip
\centerline{\includegraphics[trim={0cm 0cm 0cm 0cm},clip,width=3.6cm]{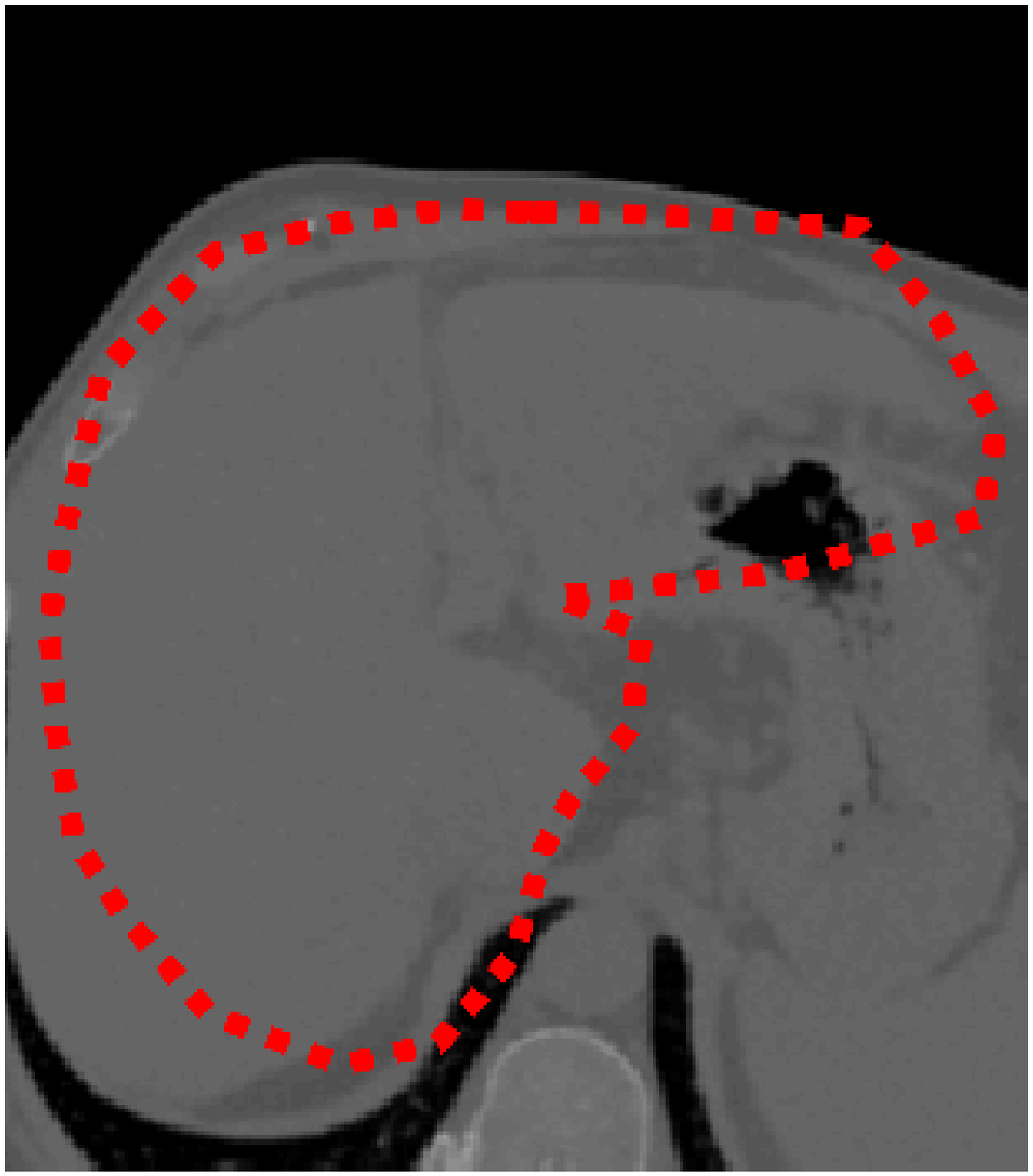}}
\centerline{(c)}\medskip
\end{minipage}
\begin{minipage}[b]{0.22\linewidth}
\centering
\centerline{\footnotesize{CT / PM-EA+Evo}}\medskip
\centerline{\includegraphics[trim={0cm 0cm 0cm 0cm},clip,width=3.6cm]{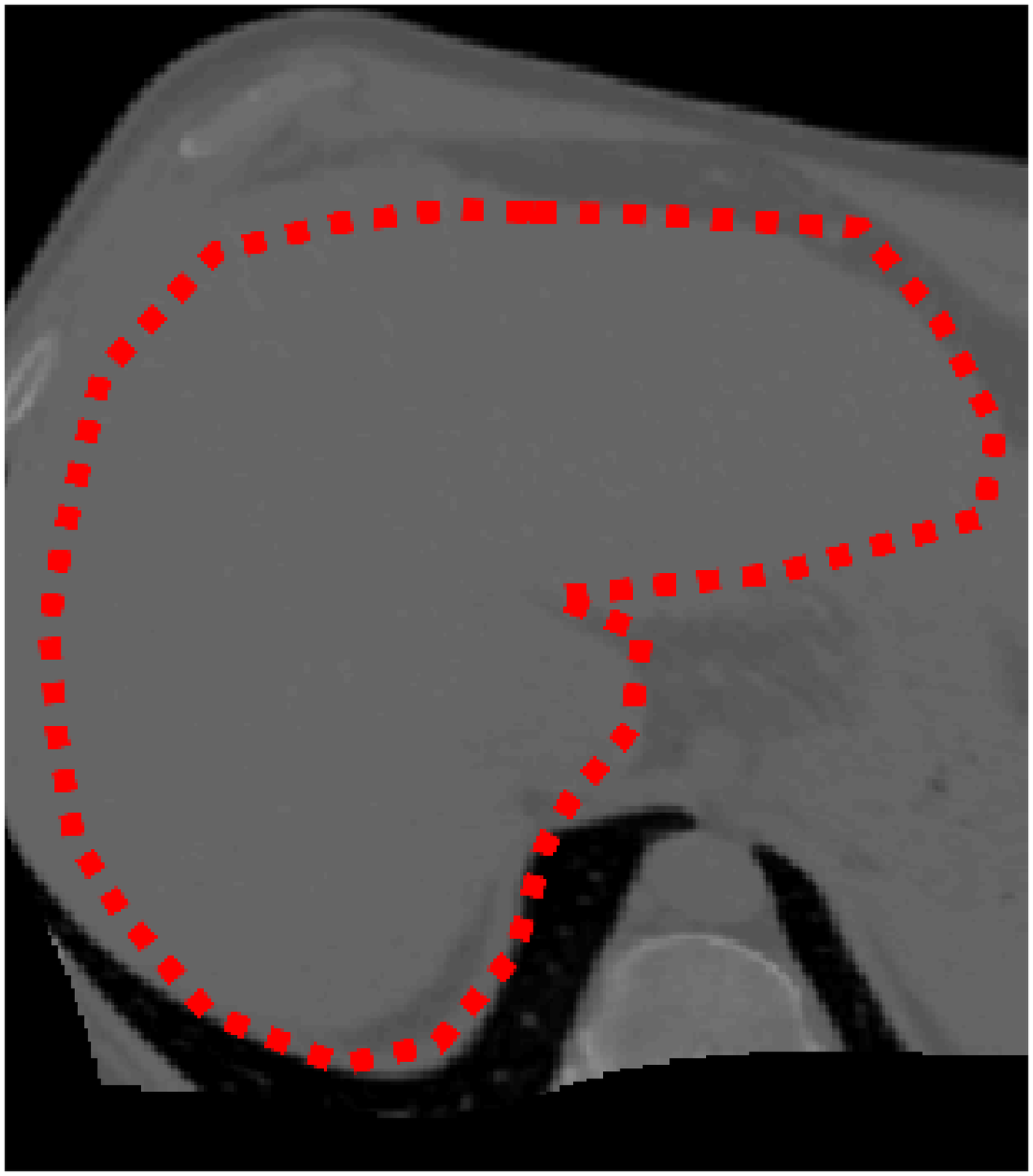}}
\centerline{(d)}\medskip
\end{minipage}

\begin{minipage}[b]{0.08\linewidth}
\centering
 \centerline{\small{Sag.}}\medskip
 \vspace{-0.2cm}
 \centerline{\small{[X-Z]}}\medskip
 \vspace{1.8cm}
\end{minipage}
\begin{minipage}[b]{0.22\linewidth}
\centering
\centerline{\includegraphics[trim={0cm 0cm 0cm 0cm},clip,width=3.6cm]{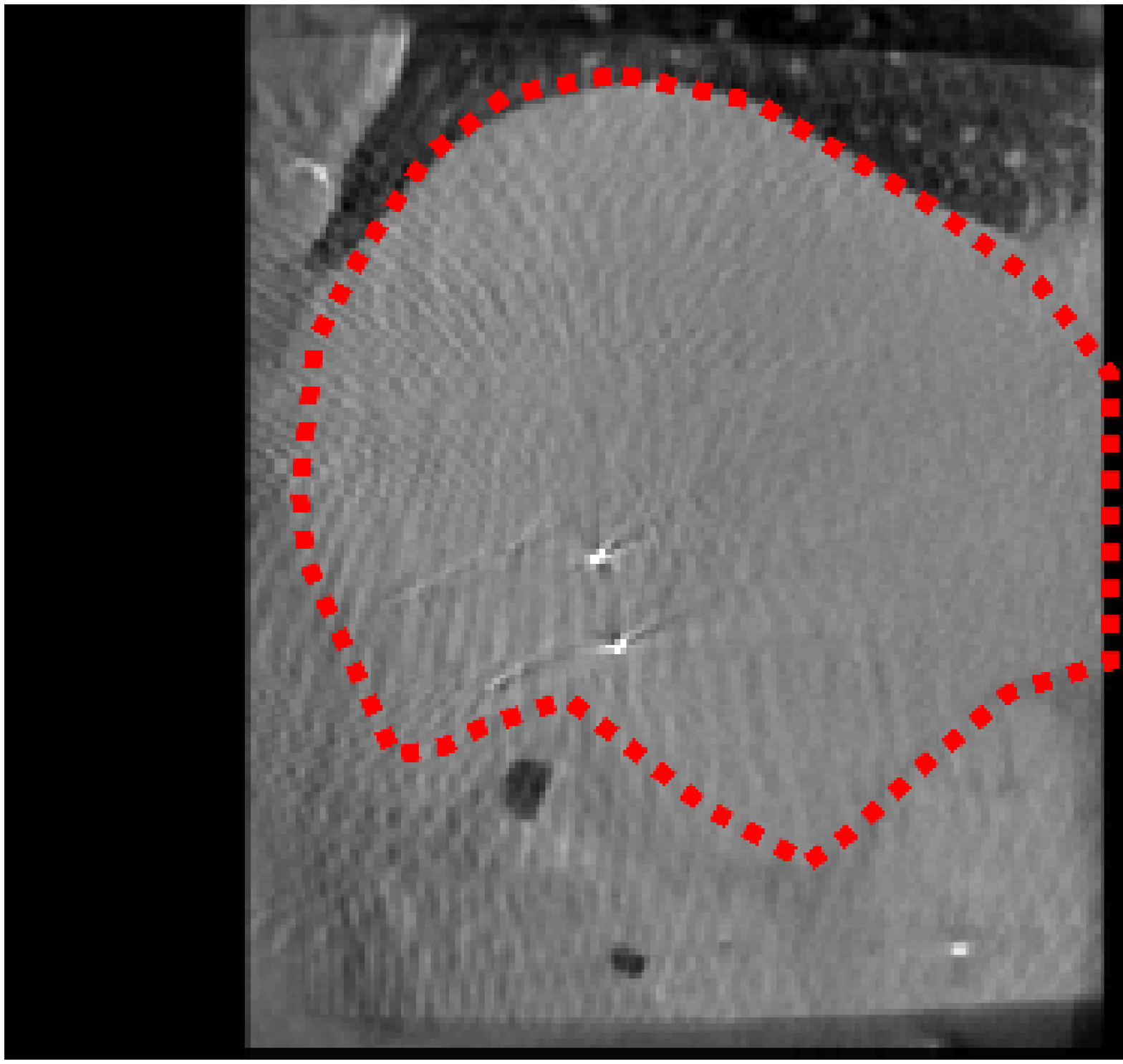}}
\centerline{(e)}\medskip
\end{minipage}
\begin{minipage}[b]{0.22\linewidth}
\centering
\centerline{\includegraphics[trim={0cm 0cm 0cm 0cm},clip,width=3.6cm]{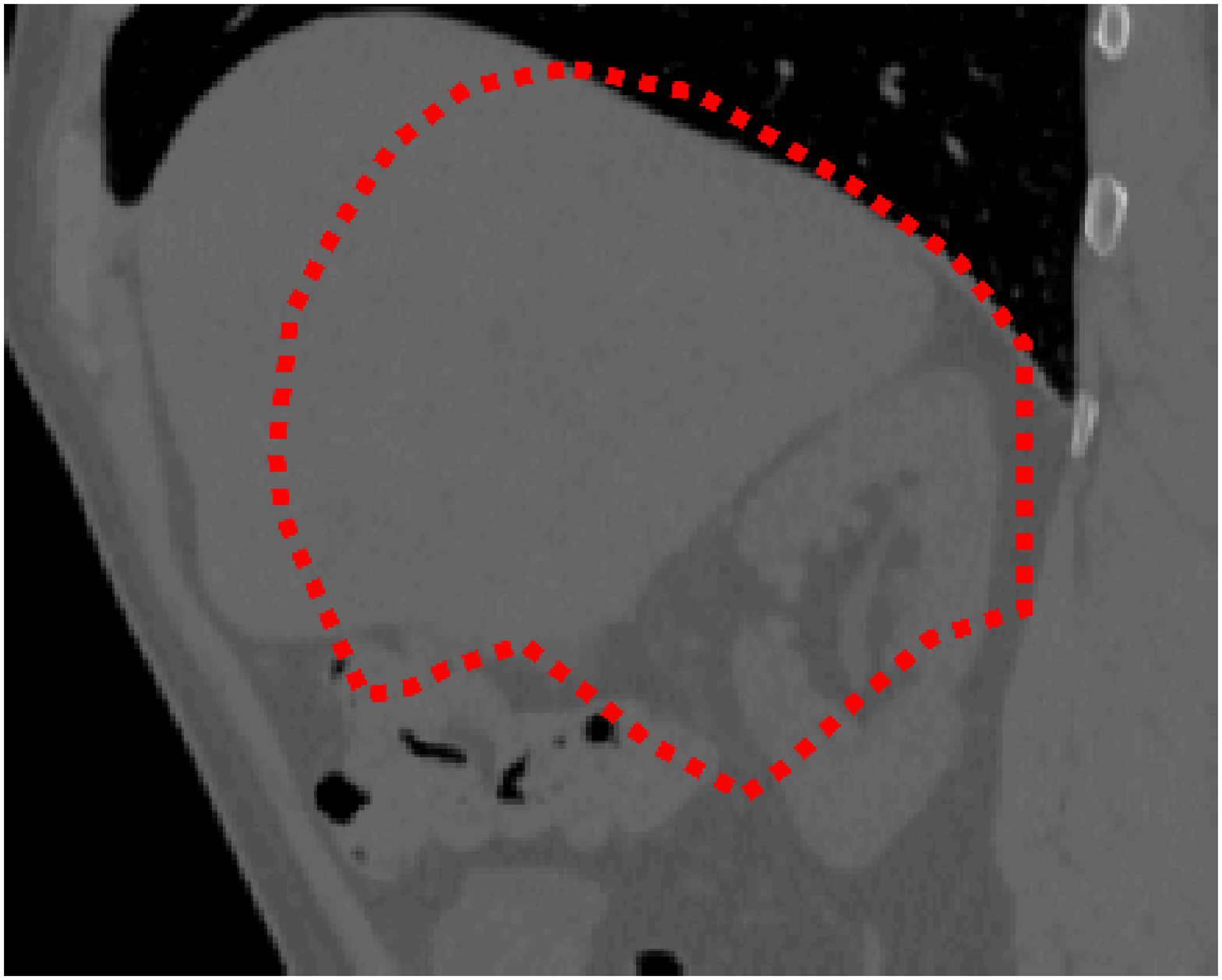}}
\centerline{(f)}\medskip
\end{minipage}
\begin{minipage}[b]{0.22\linewidth}
\centering
\centerline{\includegraphics[trim={0cm 0cm 0cm 0cm},clip,width=3.6cm]{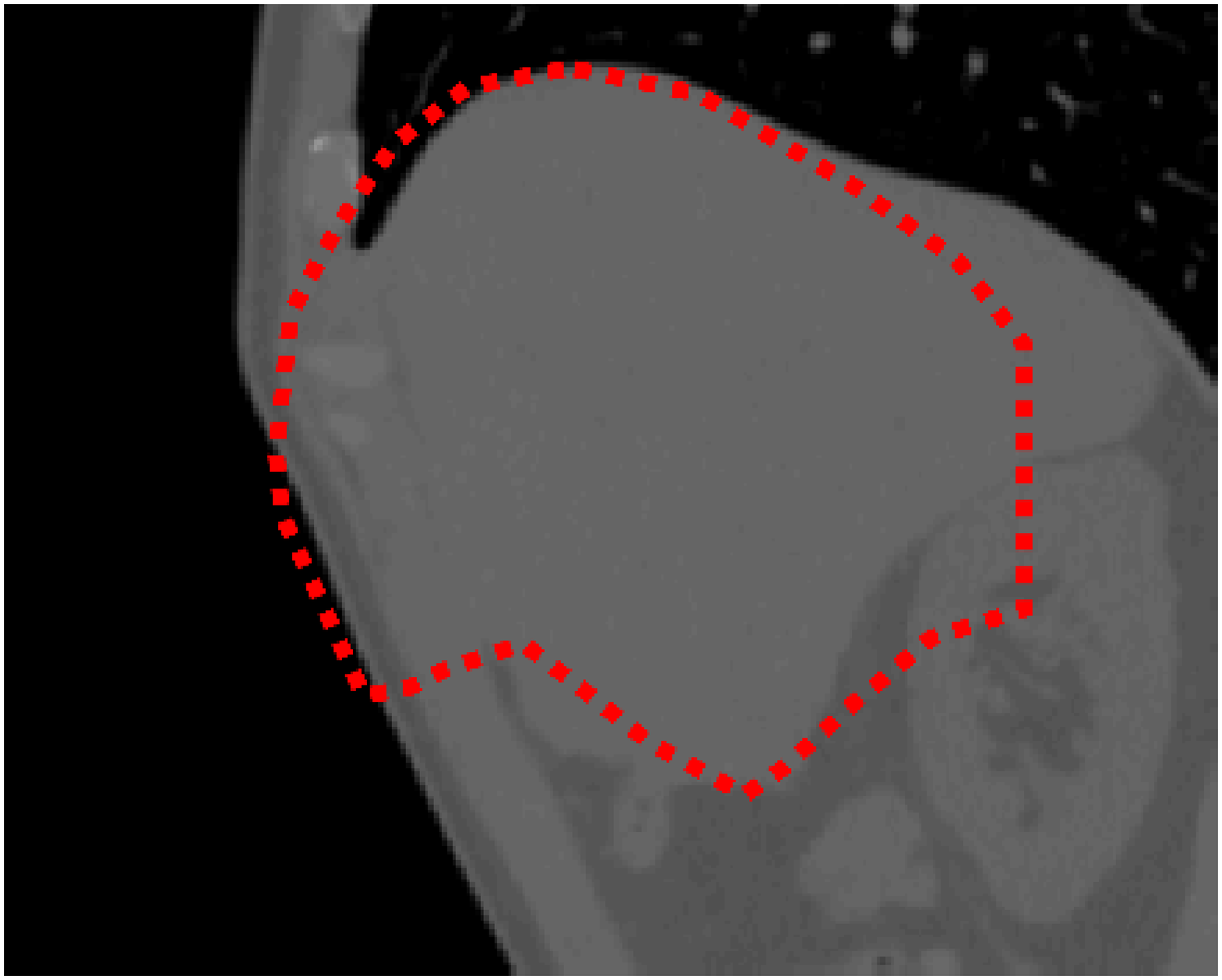}}
\centerline{(g)}\medskip
\end{minipage}
\begin{minipage}[b]{0.22\linewidth}
\centering
\centerline{\includegraphics[trim={0cm 0cm 0cm 0cm},clip,width=3.6cm]{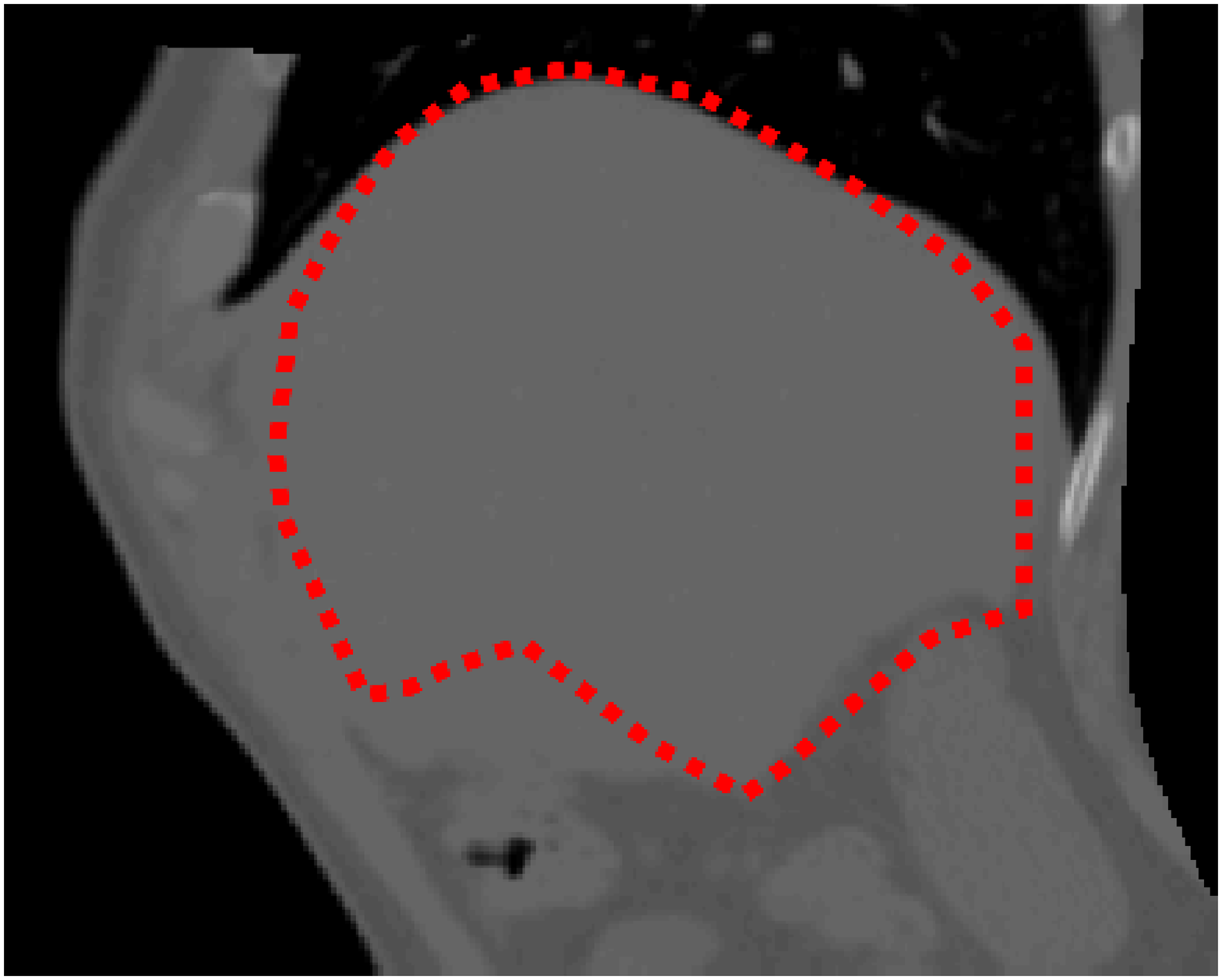}}
\centerline{(h)}\medskip
\end{minipage}

\begin{minipage}[b]{0.08\linewidth}
\centering
 \centerline{\small{Cor.}}\medskip
 \vspace{-0.2cm}
 \centerline{\small{[Y-Z]}}\medskip
 \vspace{2cm}
\end{minipage}
\begin{minipage}[b]{0.22\linewidth}
\centering
\centerline{\includegraphics[trim={0cm 0cm 0cm 0cm},clip,width=3.6cm]{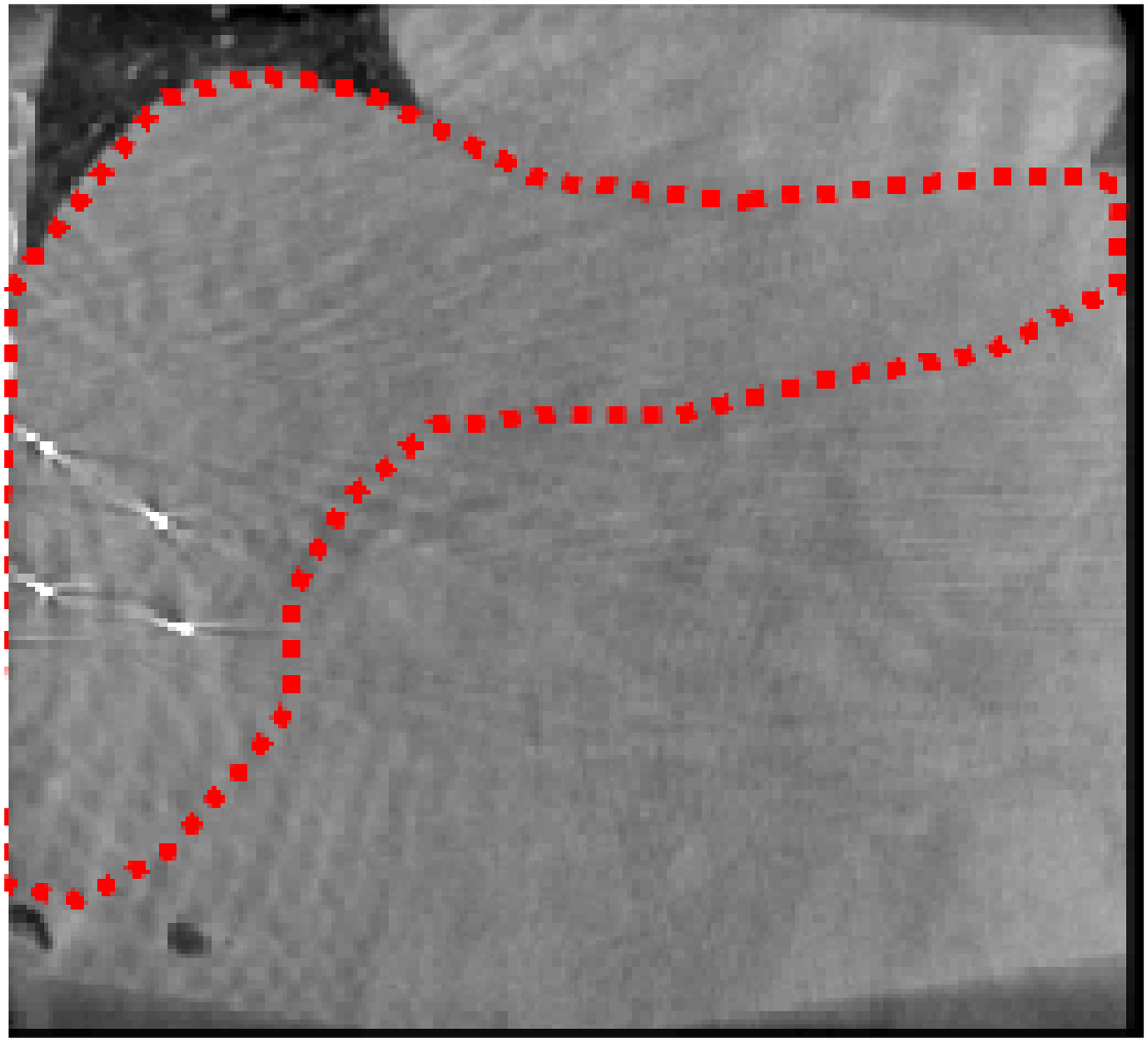}}
\centerline{(i)}\medskip
\end{minipage}
\begin{minipage}[b]{0.22\linewidth}
\centering
\centerline{\includegraphics[trim={0cm 0cm 0cm 0cm},clip,width=3.6cm]{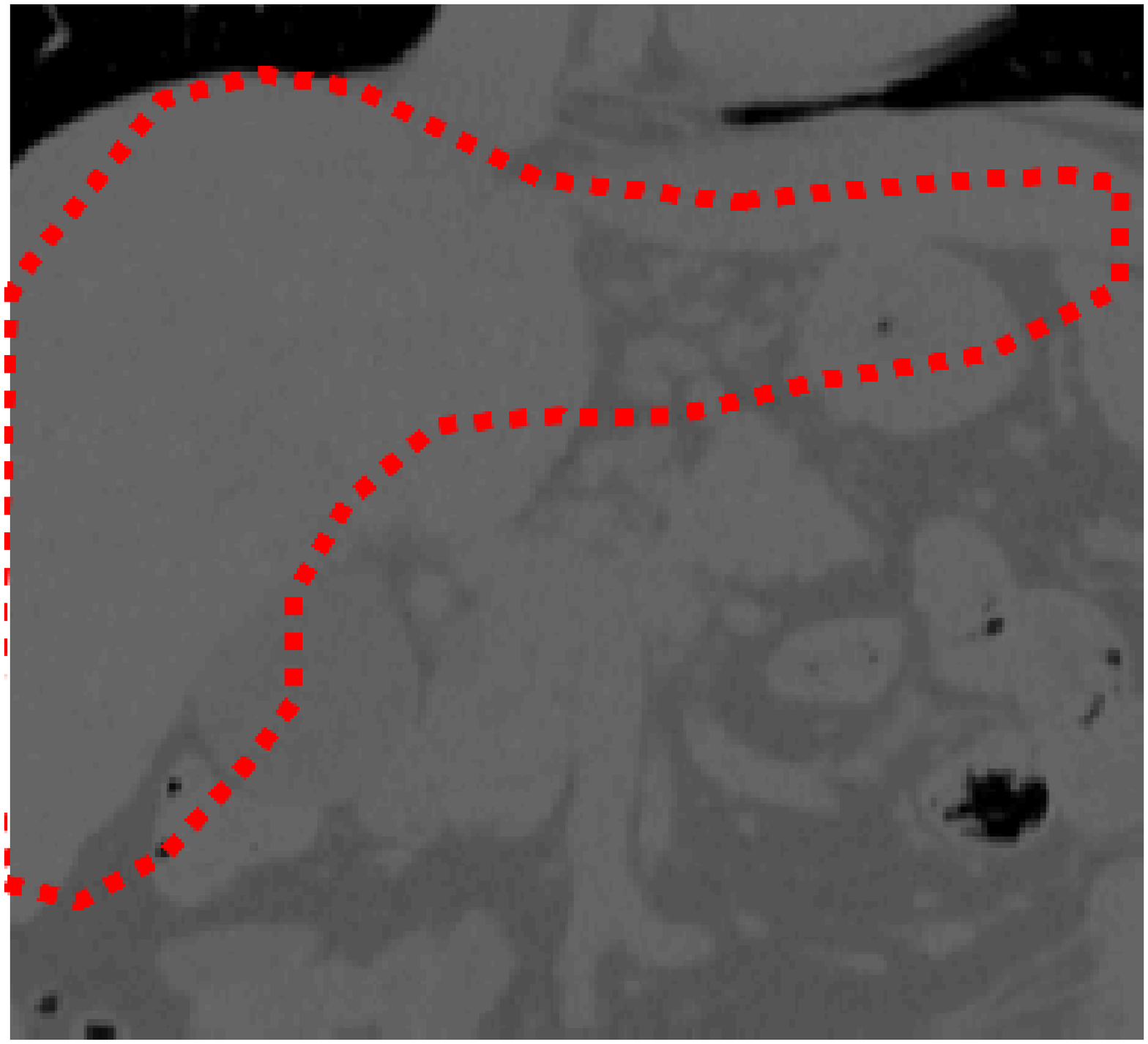}}
\centerline{(j)}\medskip
\end{minipage}
\begin{minipage}[b]{0.22\linewidth}
\centering
\centerline{\includegraphics[trim={0cm 0cm 0cm 0cm},clip,width=3.6cm]{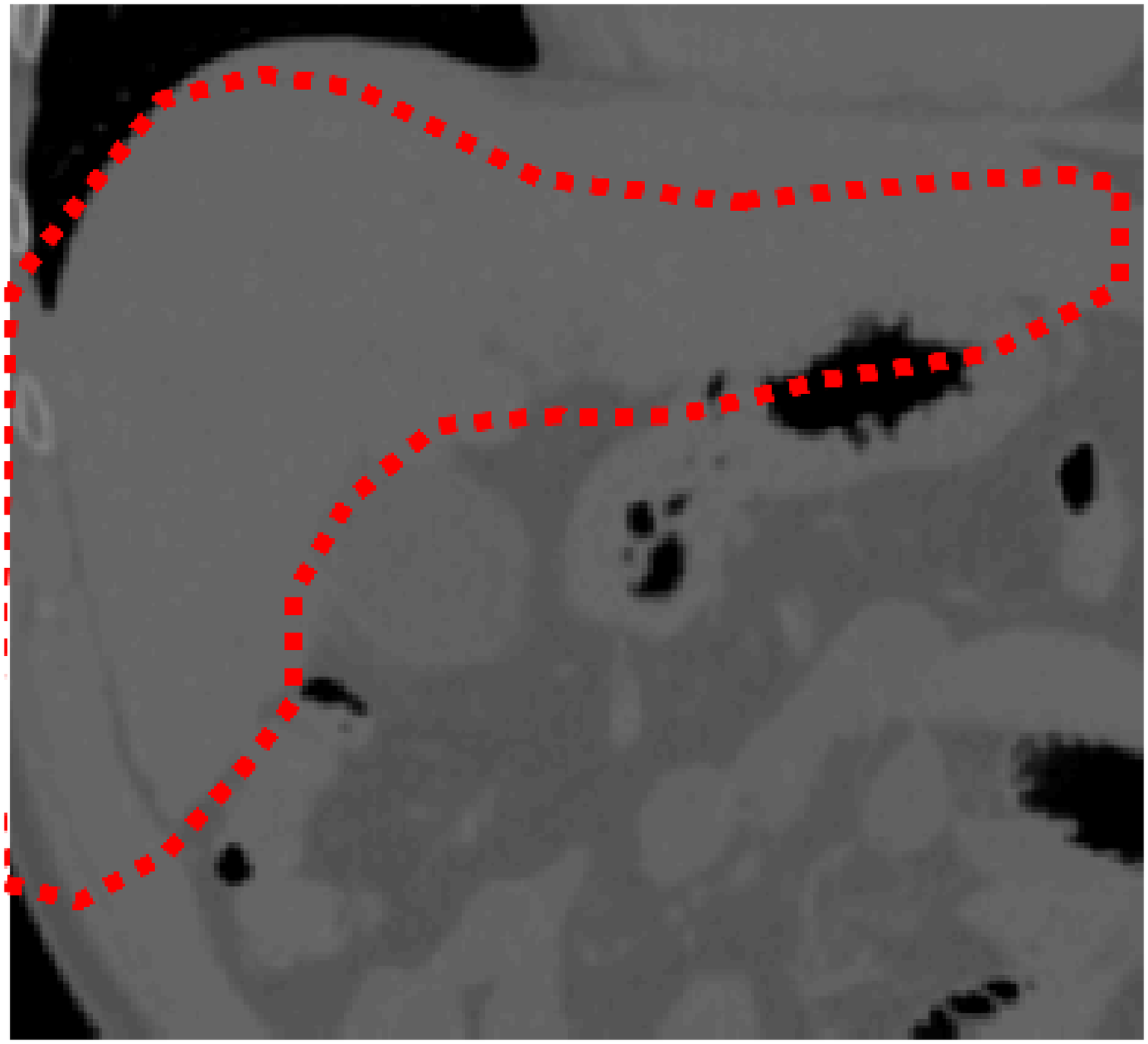}}
\centerline{(k)}\medskip
\end{minipage}
\begin{minipage}[b]{0.22\linewidth}
\centering
\centerline{\includegraphics[trim={0cm 0cm 0cm 0cm},clip,width=3.6cm]{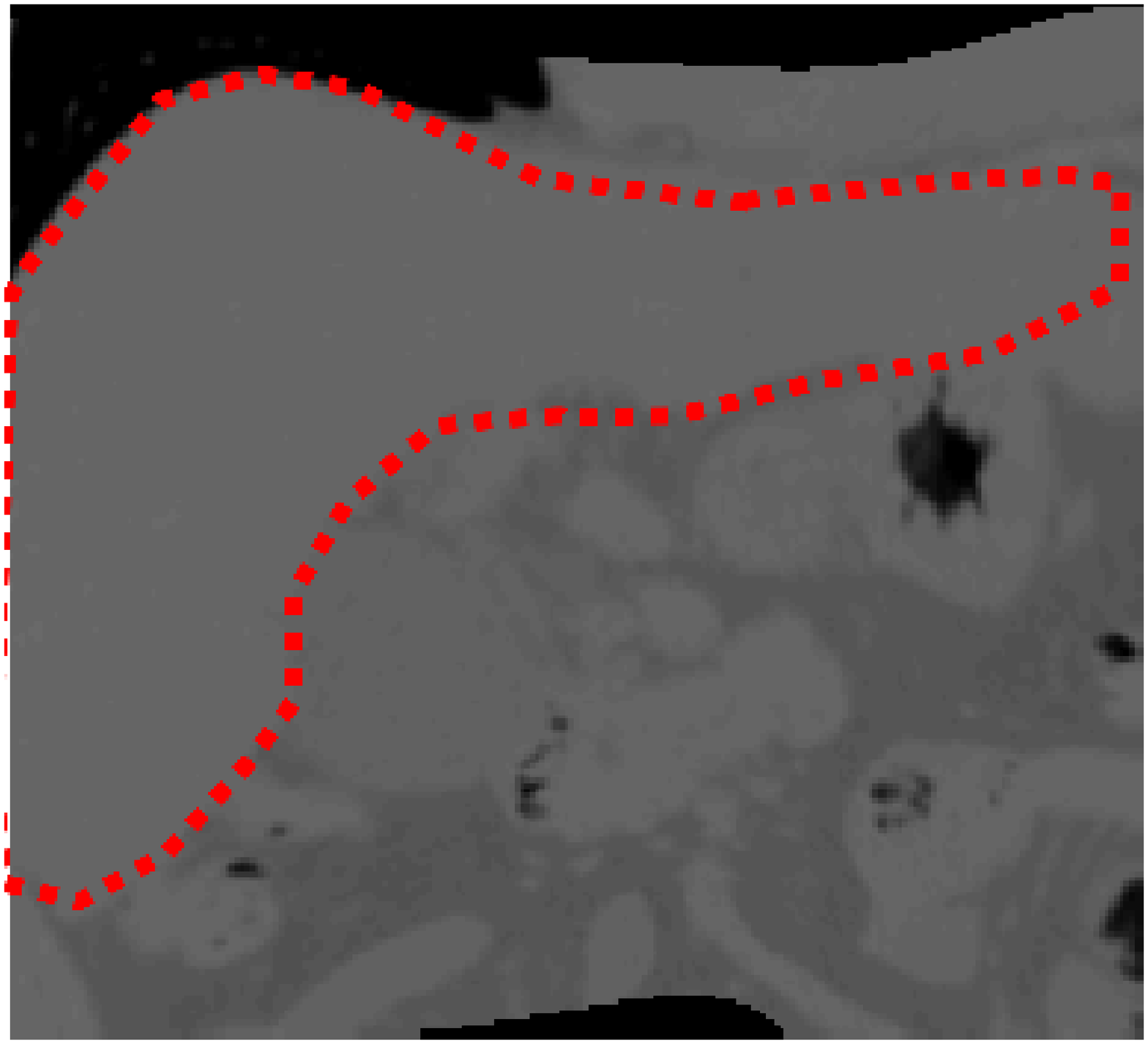}}
\centerline{(l)}\medskip
\end{minipage}

\begin{minipage}[b]{0.32\linewidth}
\centering
\centerline{\includegraphics[trim={0cm 0cm 0cm 0cm},clip,height=4.2cm]{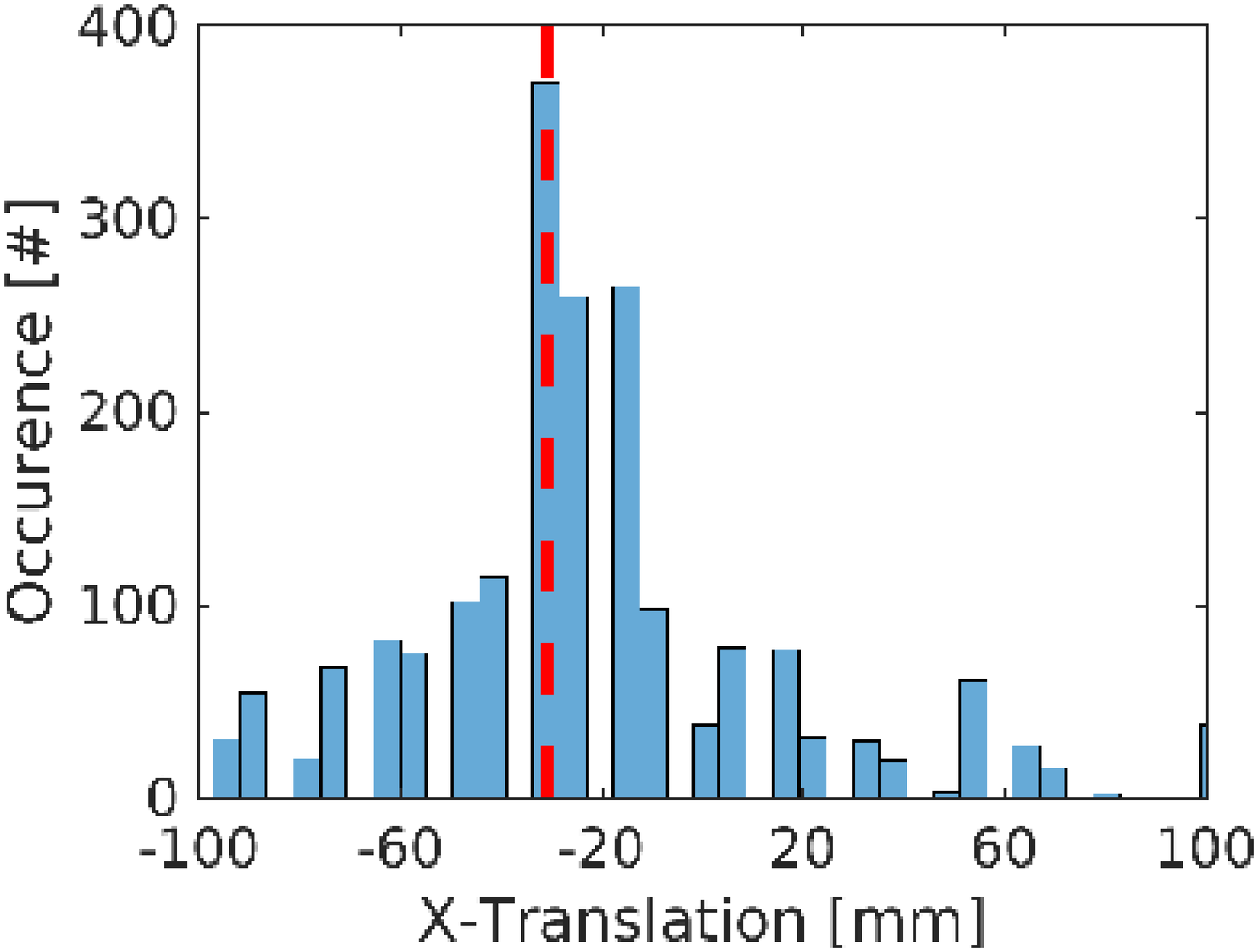}}
\centerline{(m)}\medskip
\end{minipage}
\begin{minipage}[b]{0.32\linewidth}
\centerline{\includegraphics[trim={0cm 0cm 0cm 0cm},clip,height=4.2cm]
{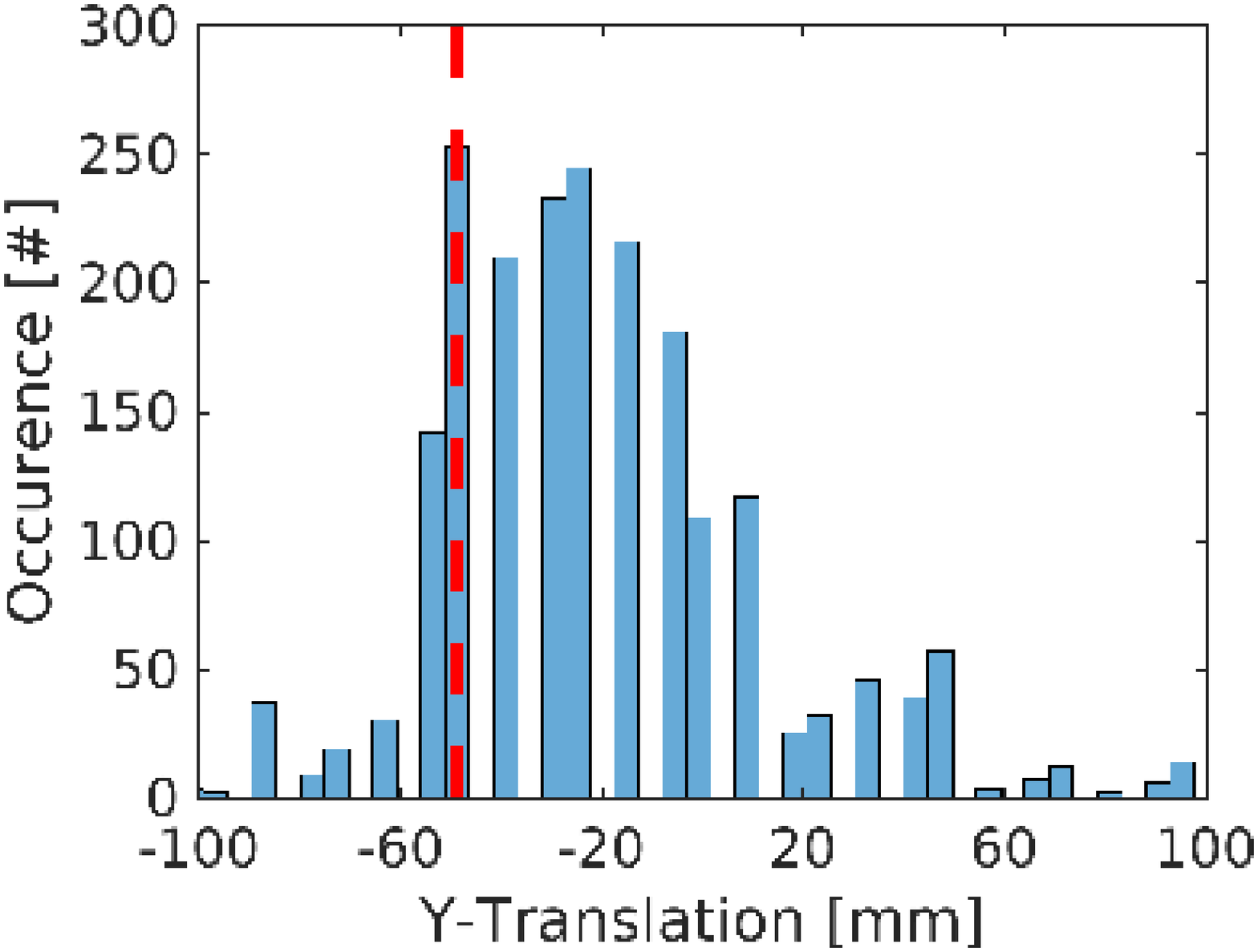}}
\centerline{(n)}\medskip
\end{minipage}
\begin{minipage}[b]{0.32\linewidth}
\centerline{\includegraphics[trim={0cm 0cm 0cm 0cm},clip,height=4.2cm]
{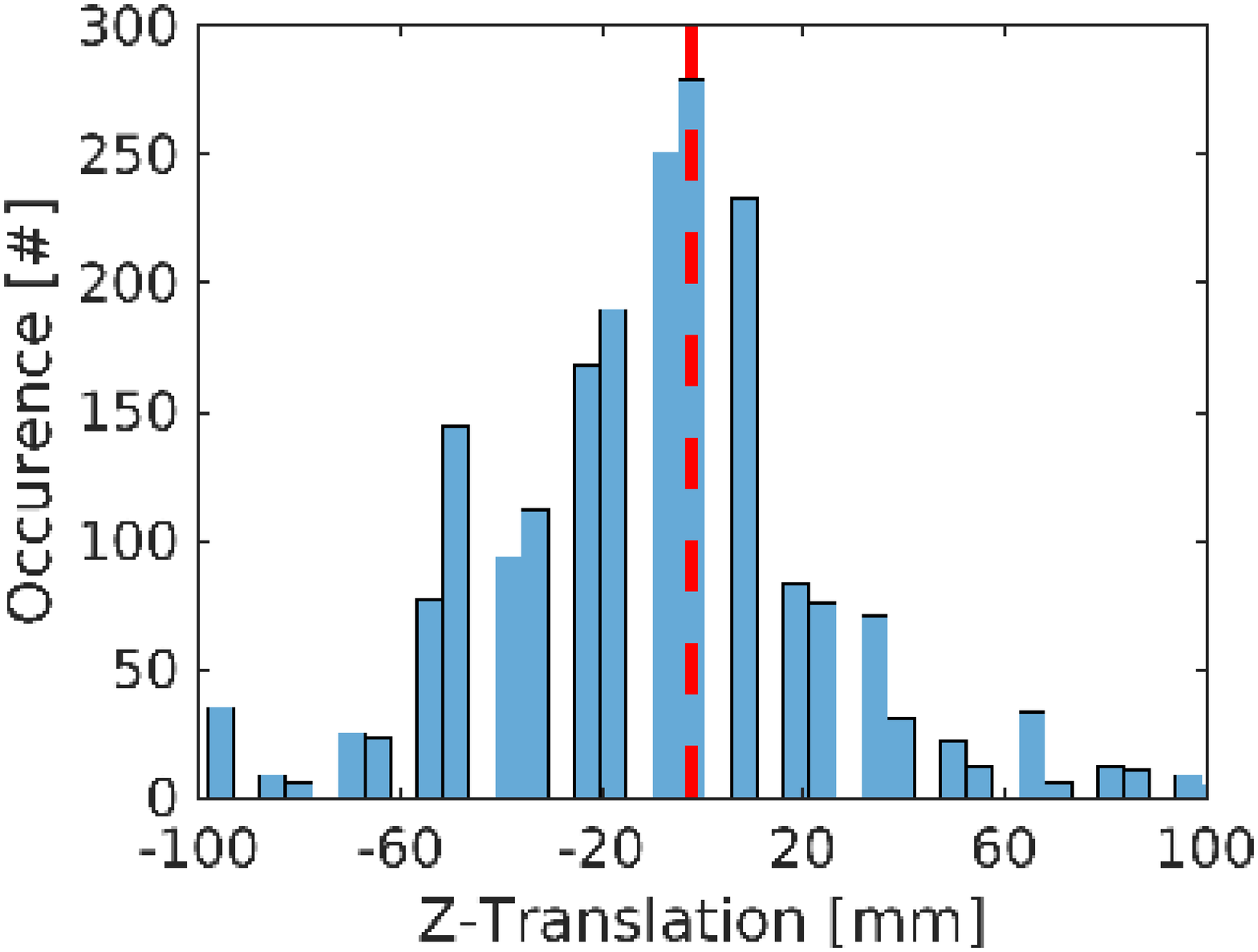}}
\centerline{(o)}\medskip
\end{minipage}

\caption{Example of a CT/CBCT registration results. The CBCT image, used as a reference for registration, was acquired immediately after insertion of 4 needles. Transversal (a-d), sagittal (e-h) and coronal (i-l) cross-sections are reported for: CBCT (first column), CT before (second column) and after registration using PM-EA (third column) and PM-EA+Evo (fourth column). Histograms of X-, Y- and Z-shifts are reported in (m), (n) and (o), respectively (maximum occurrence in red dashed line).}
\label{fig:CT_images}
\end{figure}

\begin{figure}[h!]
\begin{minipage}[b]{0.08\linewidth}
\centering
 \centerline{\small{Trans.}}\medskip
 \vspace{-0.2cm}
 \centerline{\small{[X-Y]}}\medskip
 \vspace{2.2cm}
\end{minipage}
\begin{minipage}[b]{0.22\linewidth}
\centering
\centerline{\footnotesize{CBCT}}\medskip
\vspace{0.08cm}
\centerline{\includegraphics[trim={0cm 0cm 0cm 0cm},clip,width=3.8cm]{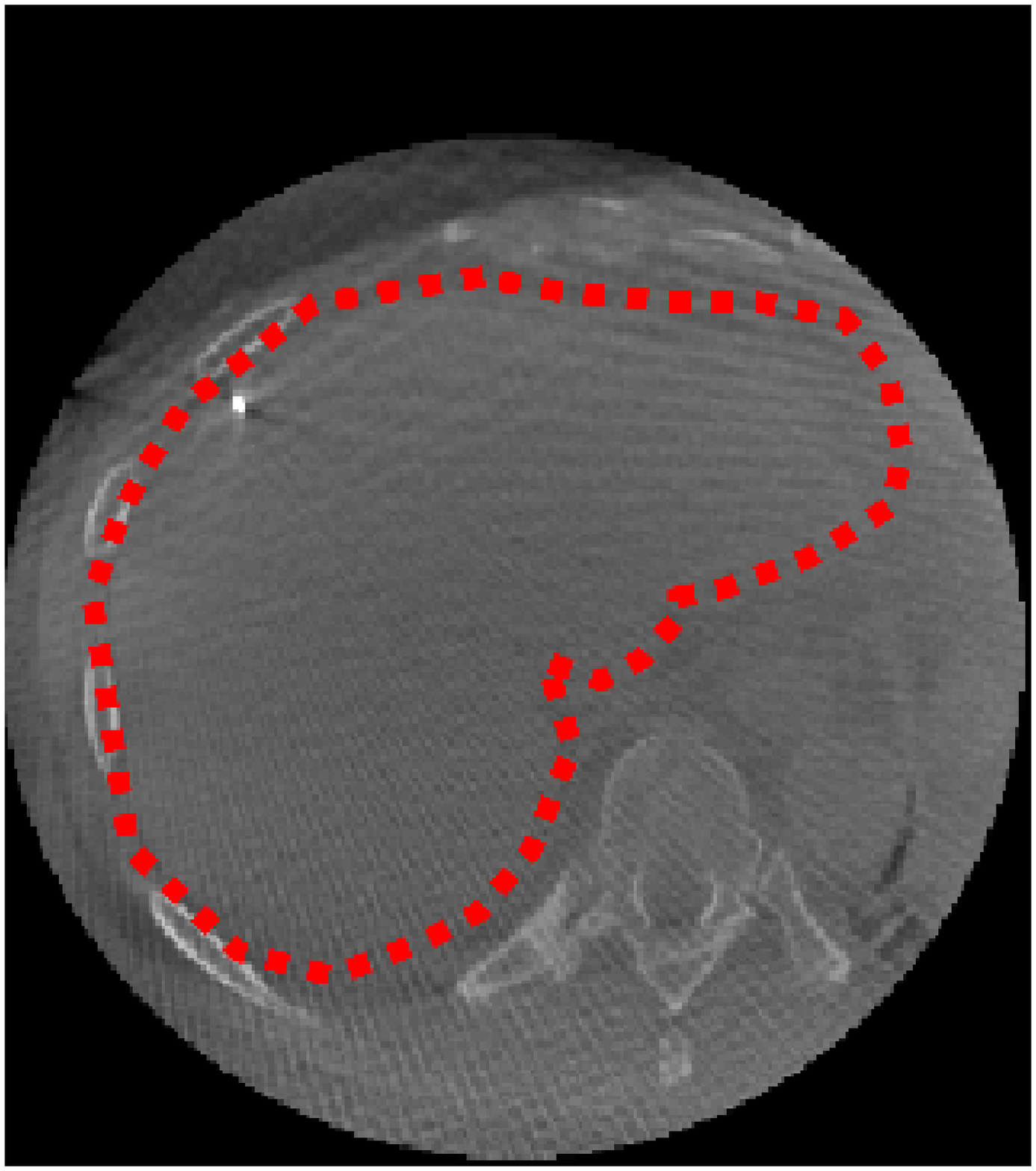}}
\centerline{(a)}\medskip
\end{minipage}
\begin{minipage}[b]{0.22\linewidth}
\centering
\centerline{\footnotesize{MRI / No registration}}\medskip
\centerline{\includegraphics[trim={0cm 0cm 0cm 0cm},clip,width=3.8cm]{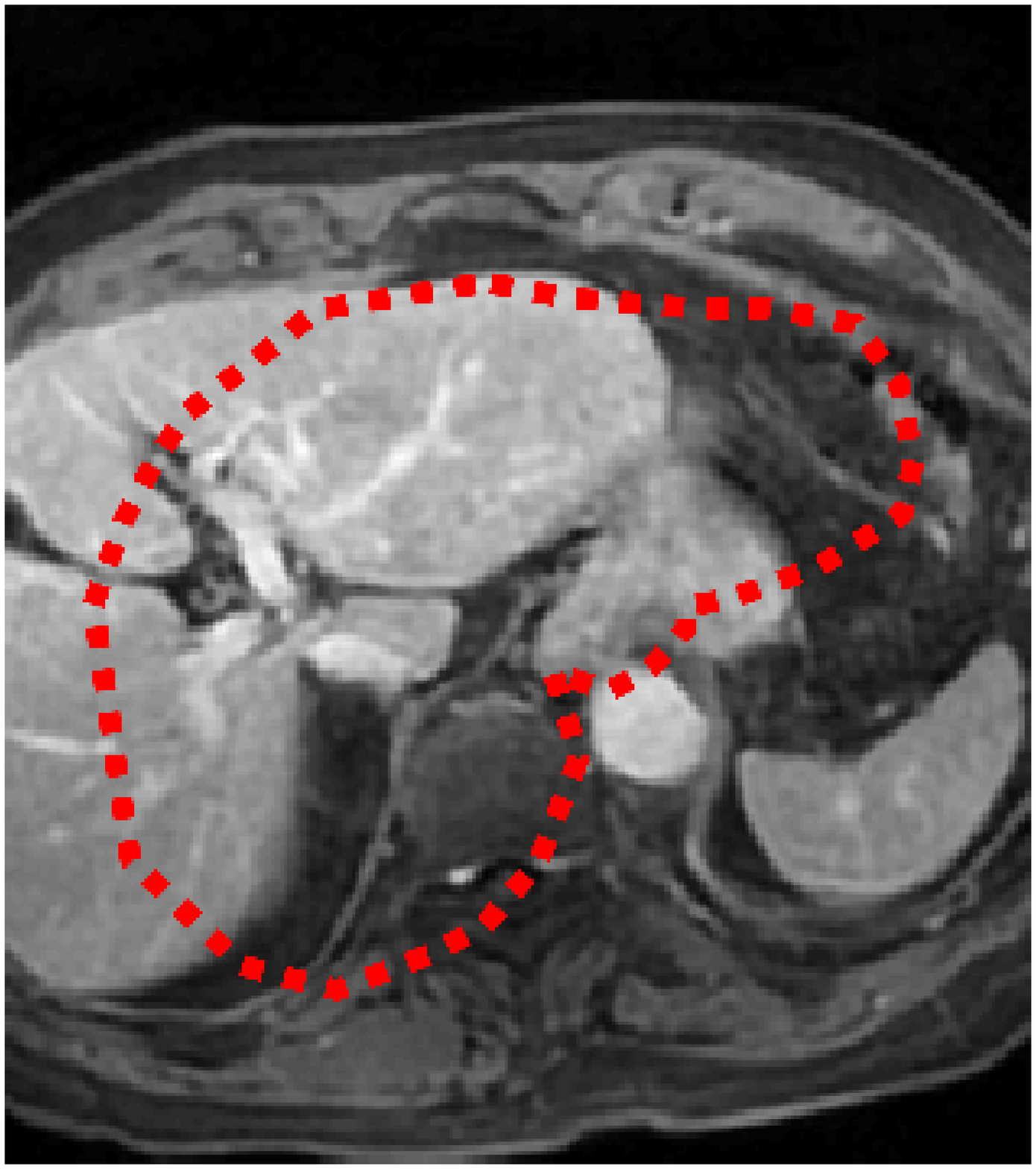}}
\centerline{(b)}\medskip
\end{minipage}
\begin{minipage}[b]{0.22\linewidth}
\centering
\centerline{\footnotesize{MRI / PM-EA}}\medskip
\centerline{\includegraphics[trim={0cm 0cm 0cm 0cm},clip,width=3.8cm]{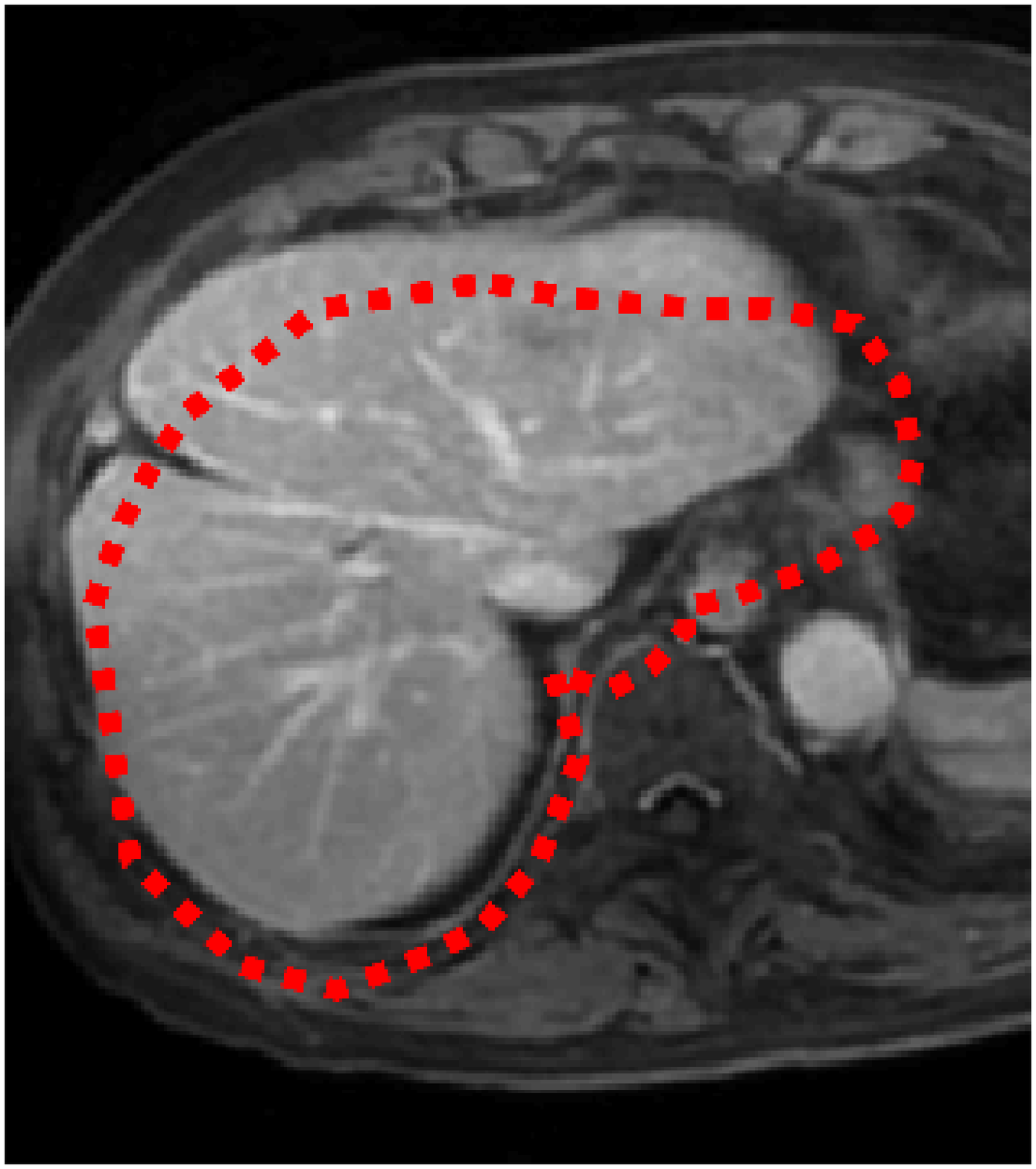}}
\centerline{(c)}\medskip
\end{minipage}
\begin{minipage}[b]{0.22\linewidth}
\centering
\centerline{\footnotesize{MRI / PM-EA+Evo}}\medskip
\centerline{\includegraphics[trim={0cm 0cm 0cm 0cm},clip,width=3.8cm]{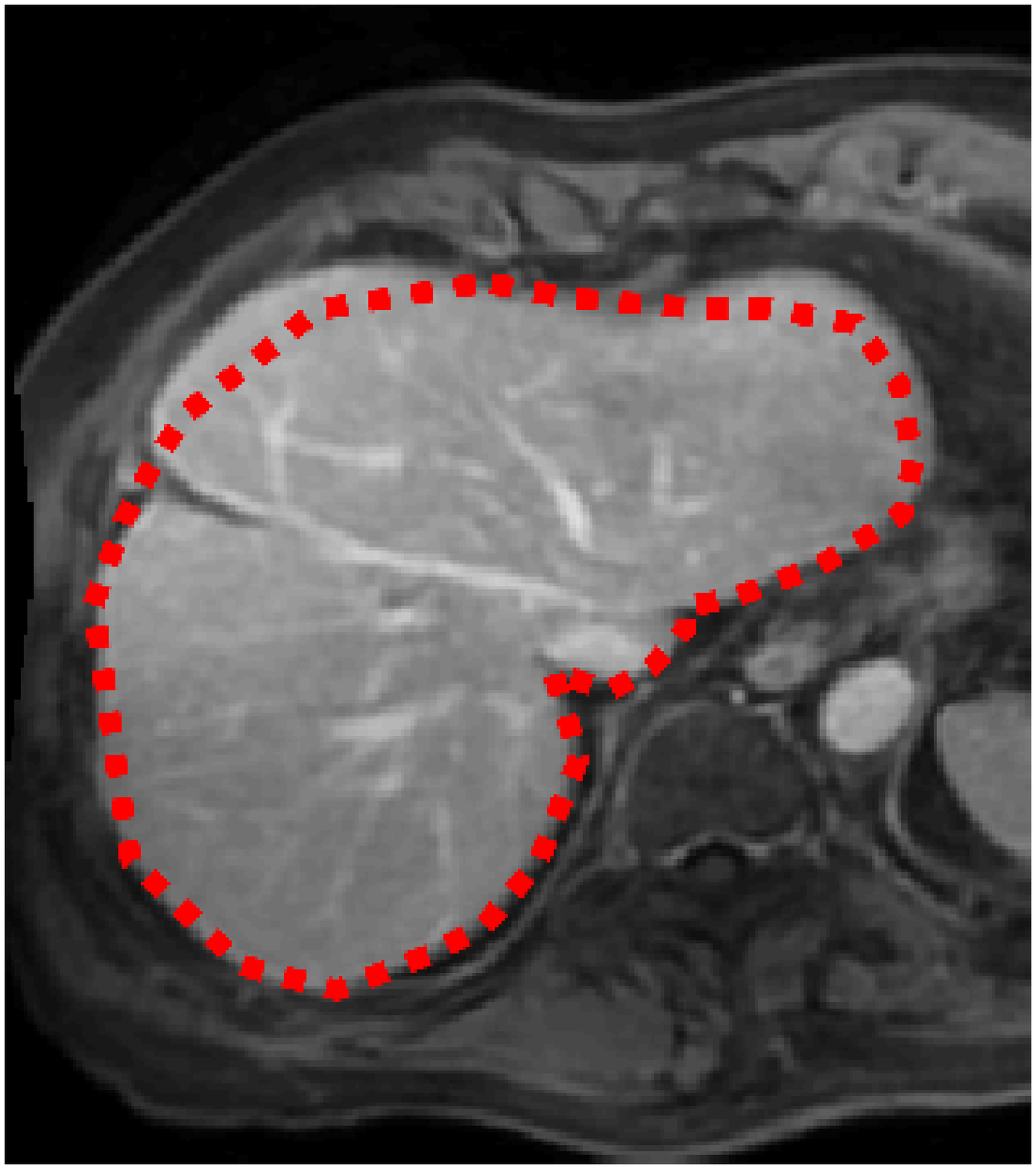}}
\centerline{(d)}\medskip
\end{minipage}

\begin{minipage}[b]{0.08\linewidth}
\centering
 \centerline{\small{Sag.}}\medskip
 \vspace{-0.2cm}
 \centerline{\small{[X-Z]}}\medskip
 \vspace{1.8cm}
\end{minipage}
\begin{minipage}[b]{0.22\linewidth}
\centering
\centerline{\includegraphics[trim={0cm 0.25cm 0cm 0.25cm},clip,width=3.8cm]{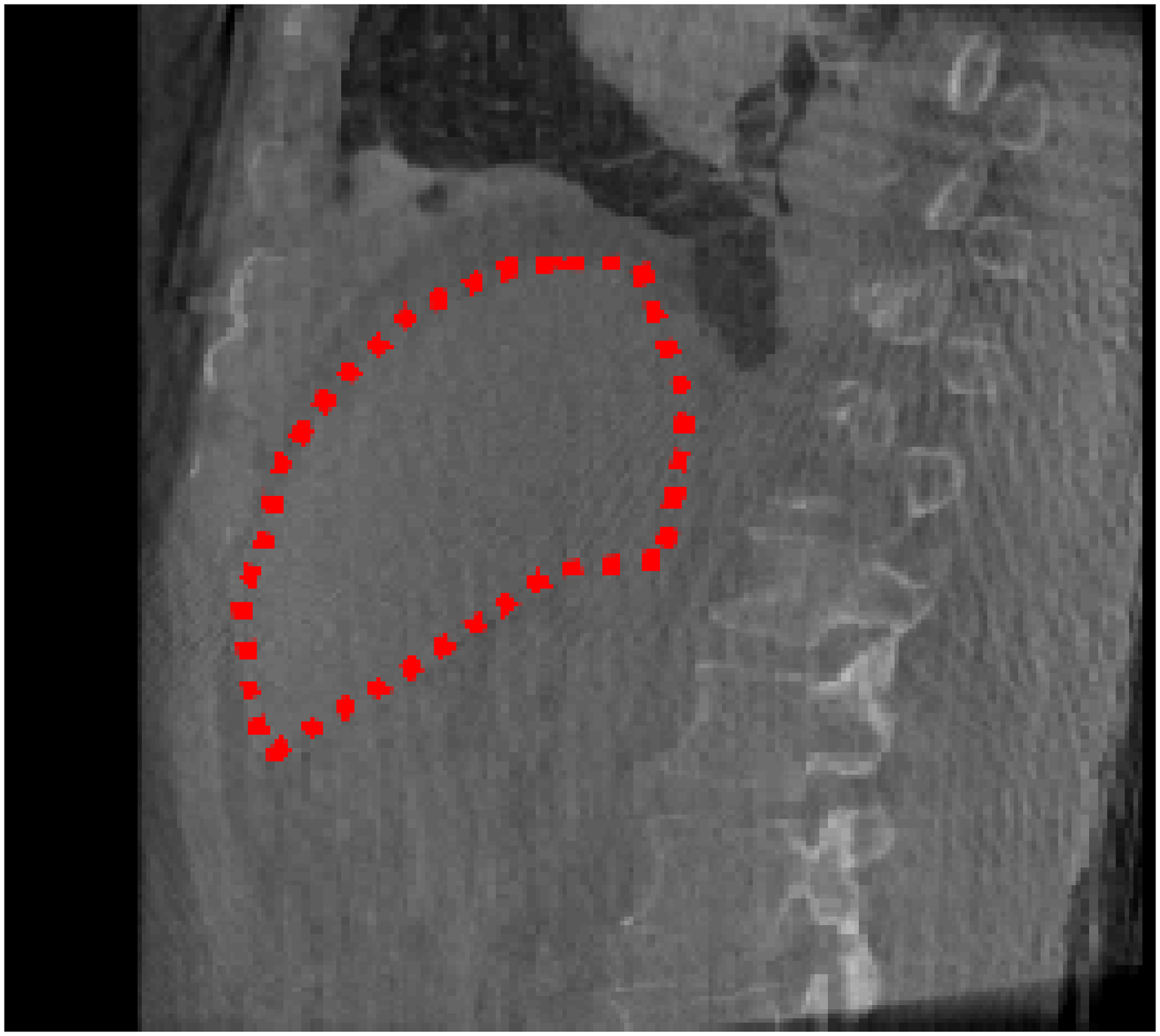}}
\centerline{(e)}\medskip
\end{minipage}
\begin{minipage}[b]{0.22\linewidth}
\centering
\centerline{\includegraphics[trim={0cm 0.25cm 0cm 0.25cm},clip,width=3.8cm]{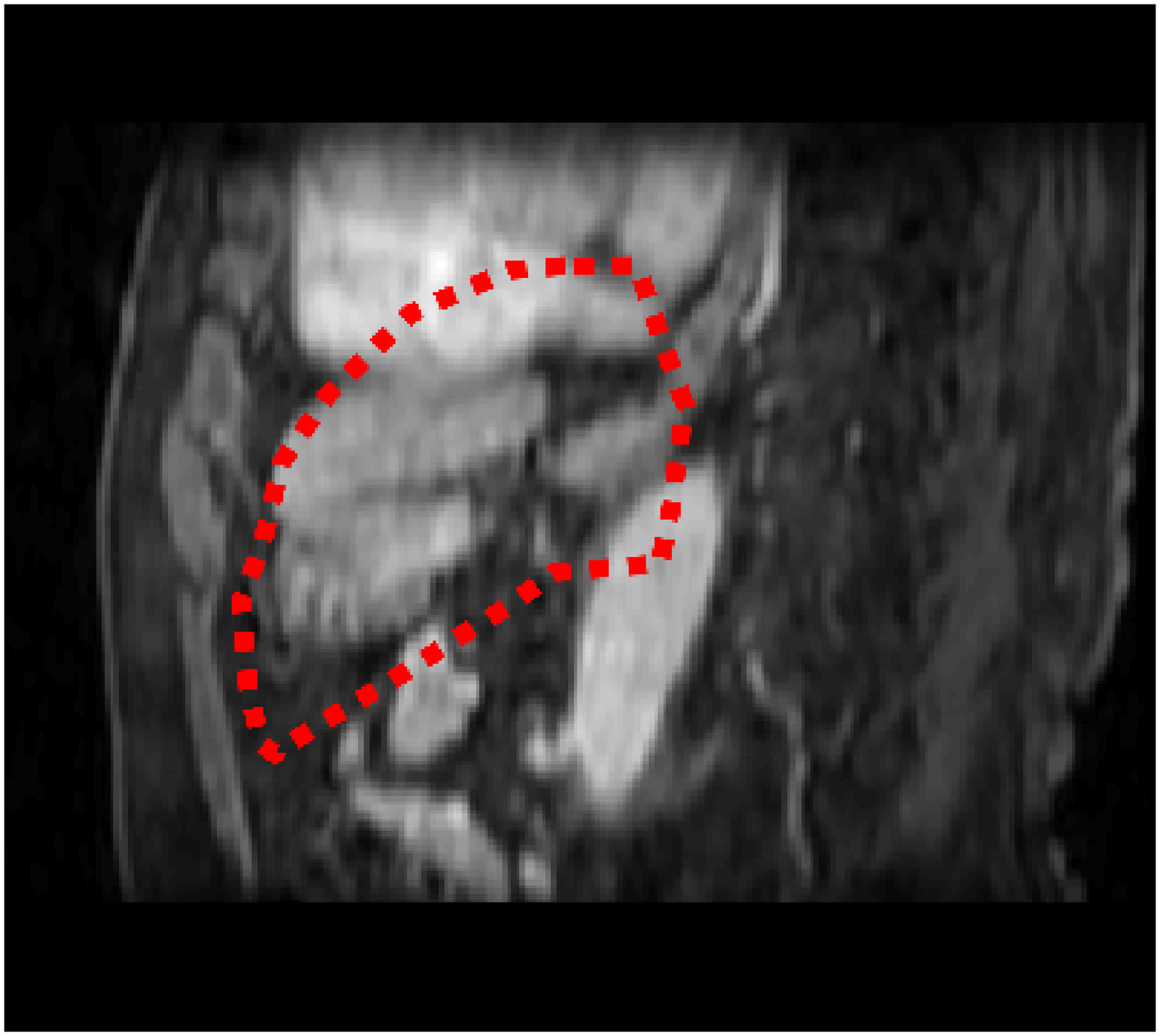}}
\centerline{(f)}\medskip
\end{minipage}
\begin{minipage}[b]{0.22\linewidth}
\centering
\centerline{\includegraphics[trim={0cm 0.25cm 0cm 0.25cm},clip,width=3.8cm]{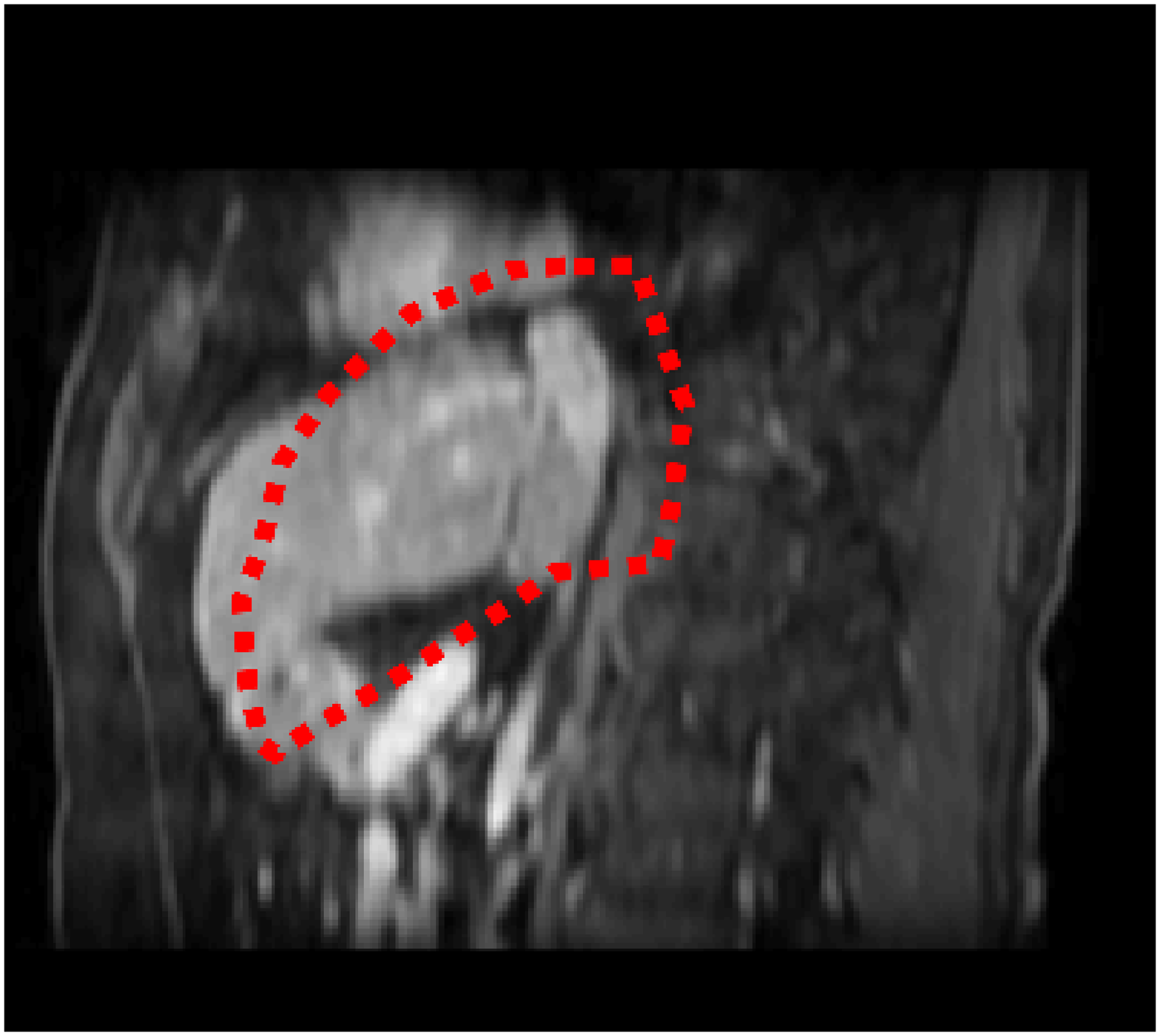}}
\centerline{(g)}\medskip
\end{minipage}
\begin{minipage}[b]{0.22\linewidth}
\centering
\centerline{\includegraphics[trim={0cm 0.25cm 0cm 0.25cm},clip,width=3.8cm]{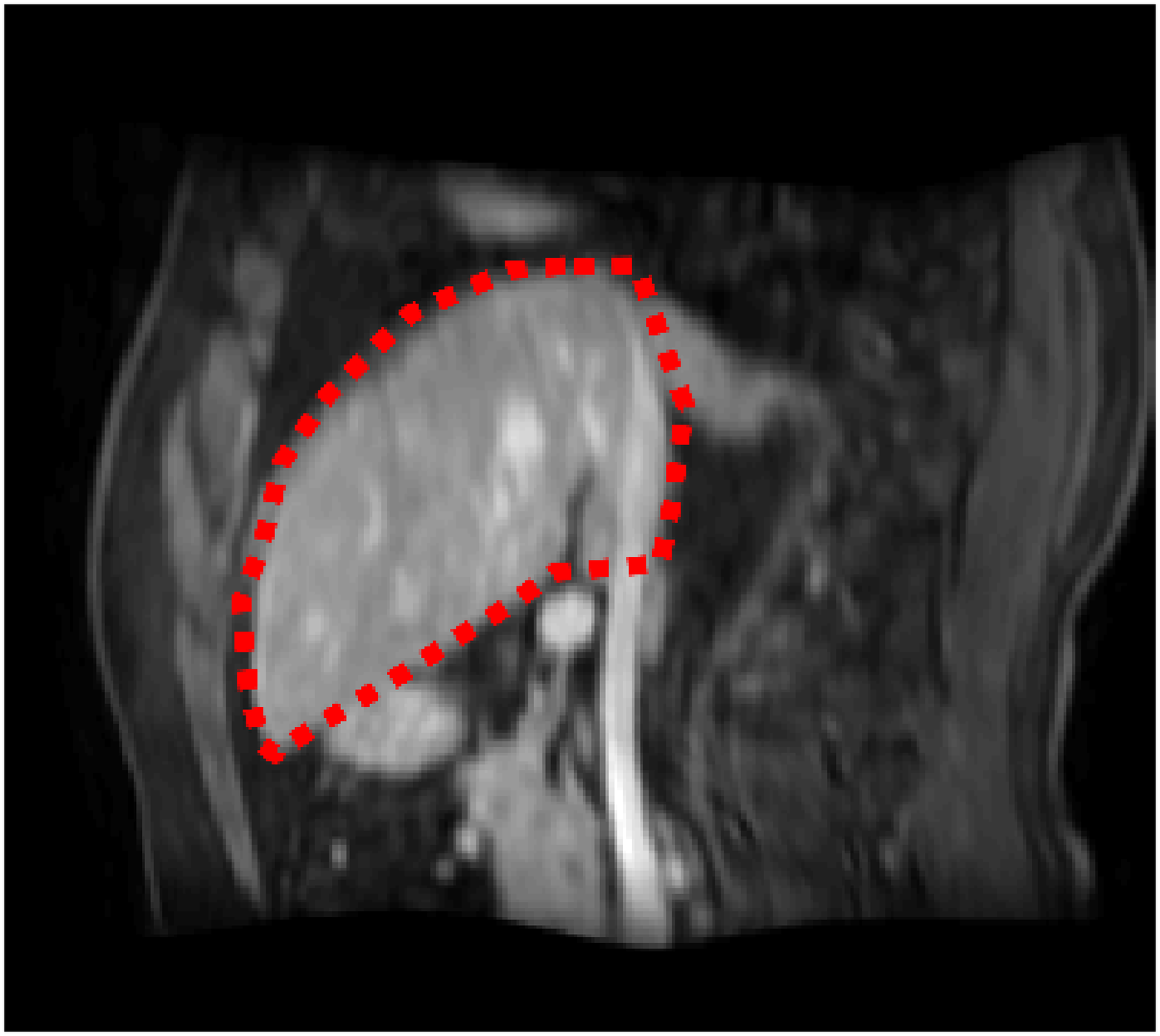}}
\centerline{(h)}\medskip
\end{minipage}

\begin{minipage}[b]{0.08\linewidth}
\centering
 \centerline{\small{Cor.}}\medskip
 \vspace{-0.2cm}
 \centerline{\small{[Y-Z]}}\medskip
 \vspace{2cm}
\end{minipage}
\begin{minipage}[b]{0.22\linewidth}
\centering
\centerline{\includegraphics[trim={0cm 1cm 0cm 1cm},clip,width=3.8cm]{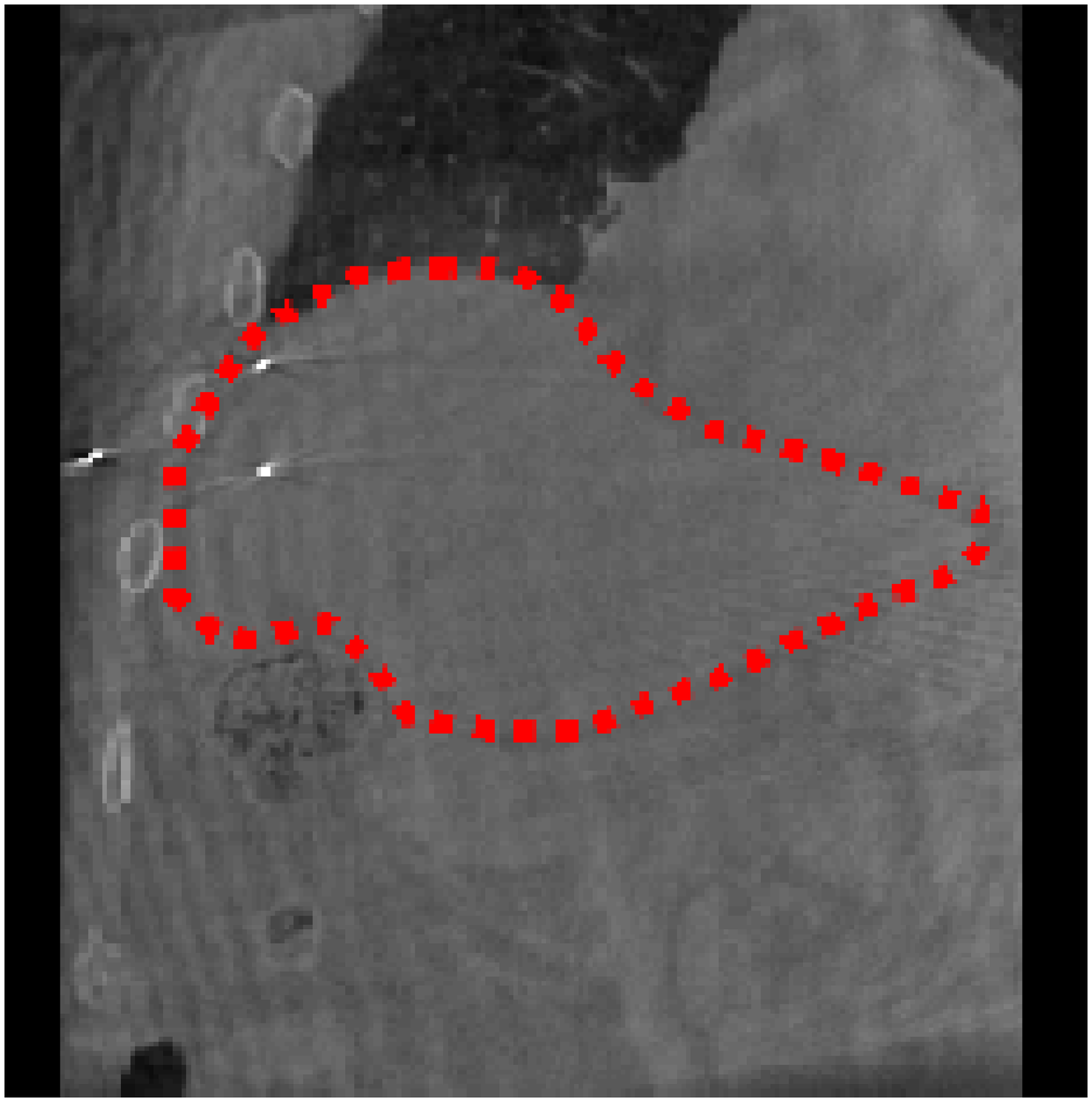}}
\centerline{(i)}\medskip
\end{minipage}
\begin{minipage}[b]{0.22\linewidth}
\centering
\centerline{\includegraphics[trim={0cm 1cm 0cm 1cm},clip,width=3.8cm]{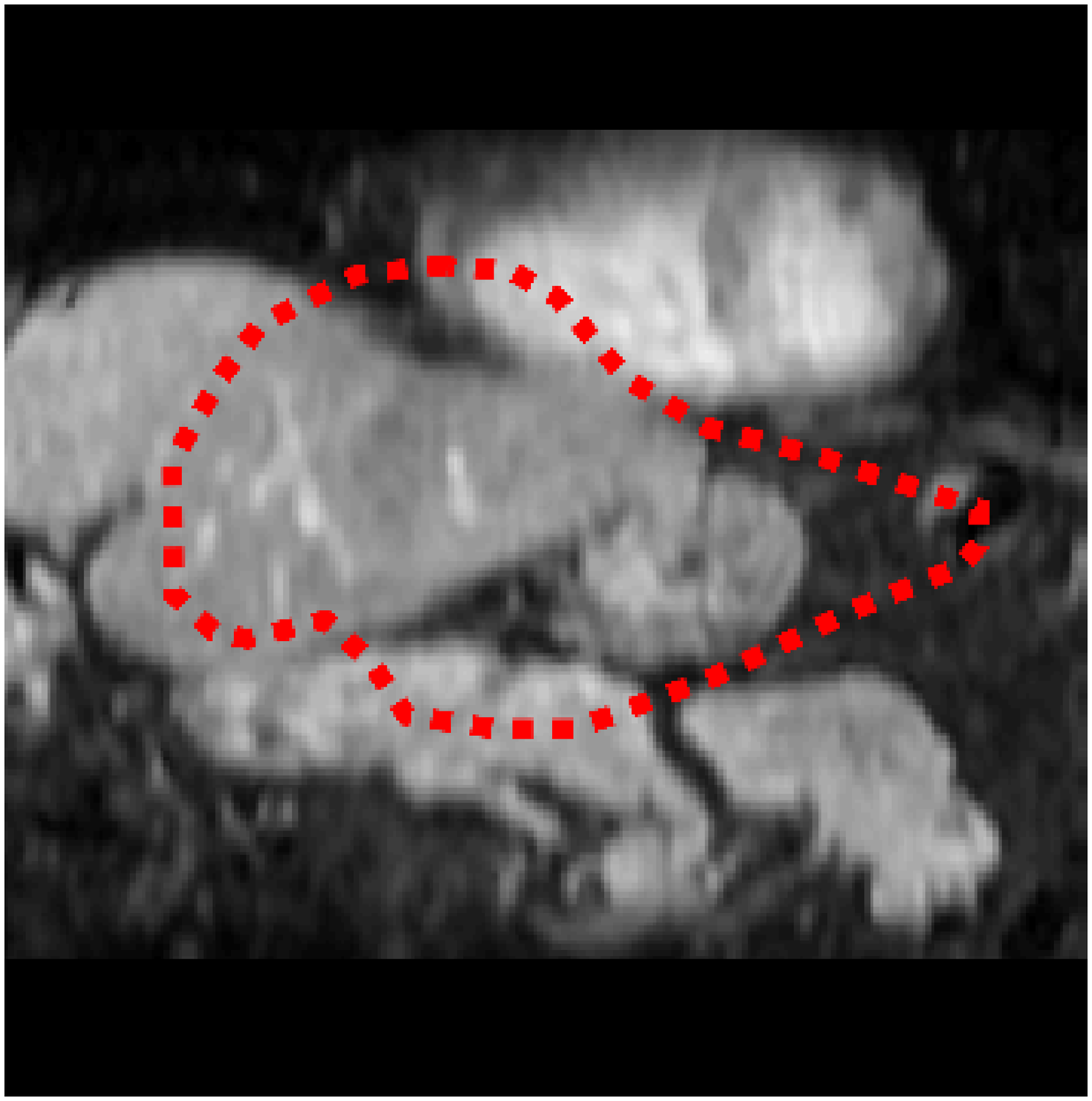}}
\centerline{(j)}\medskip
\end{minipage}
\begin{minipage}[b]{0.22\linewidth}
\centering
\centerline{\includegraphics[trim={0cm 1cm 0cm 1cm},clip,width=3.8cm]{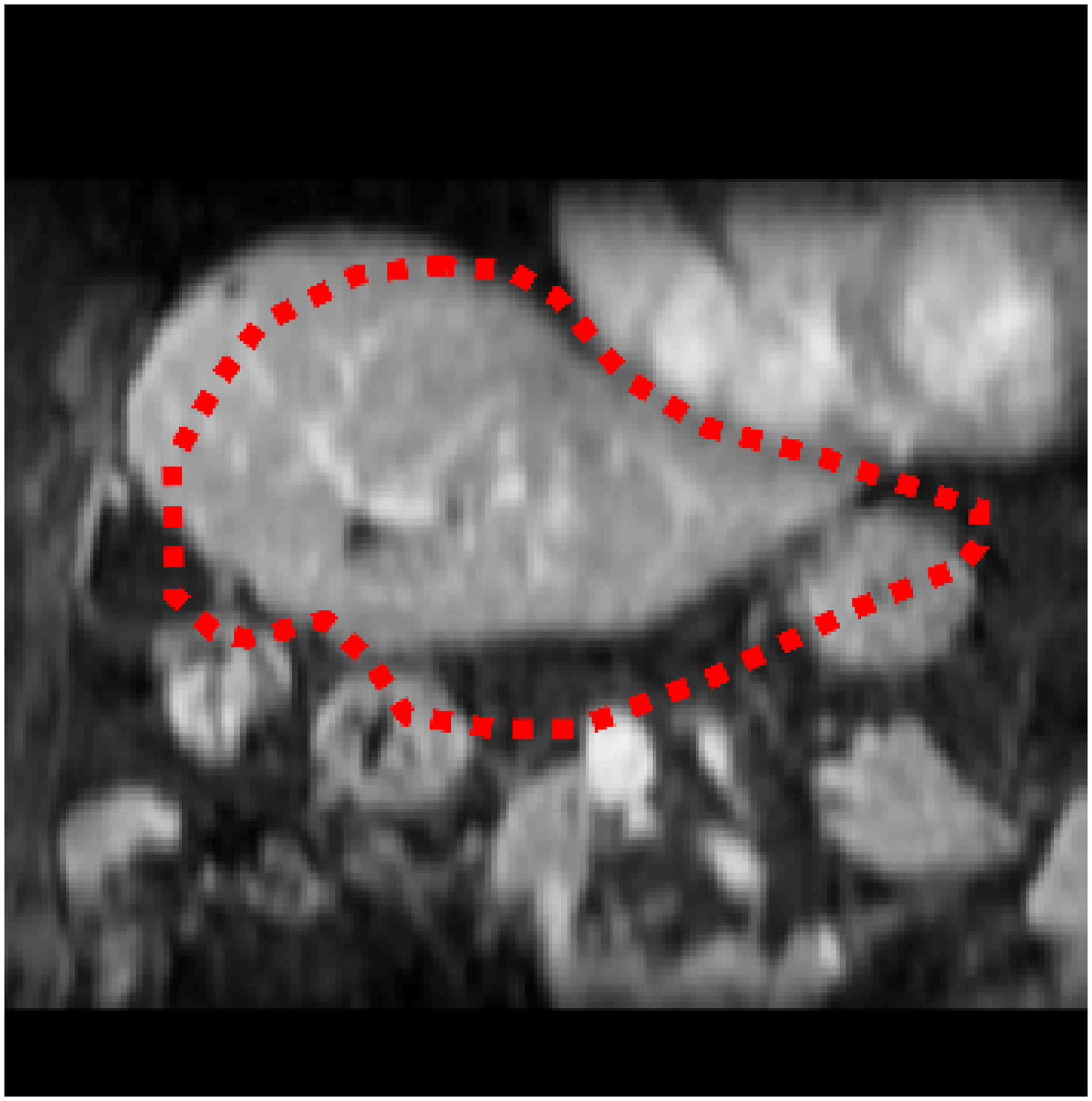}}
\centerline{(k)}\medskip
\end{minipage}
\begin{minipage}[b]{0.22\linewidth}
\centering
\centerline{\includegraphics[trim={0cm 1cm 0cm 1cm},clip,width=3.8cm]{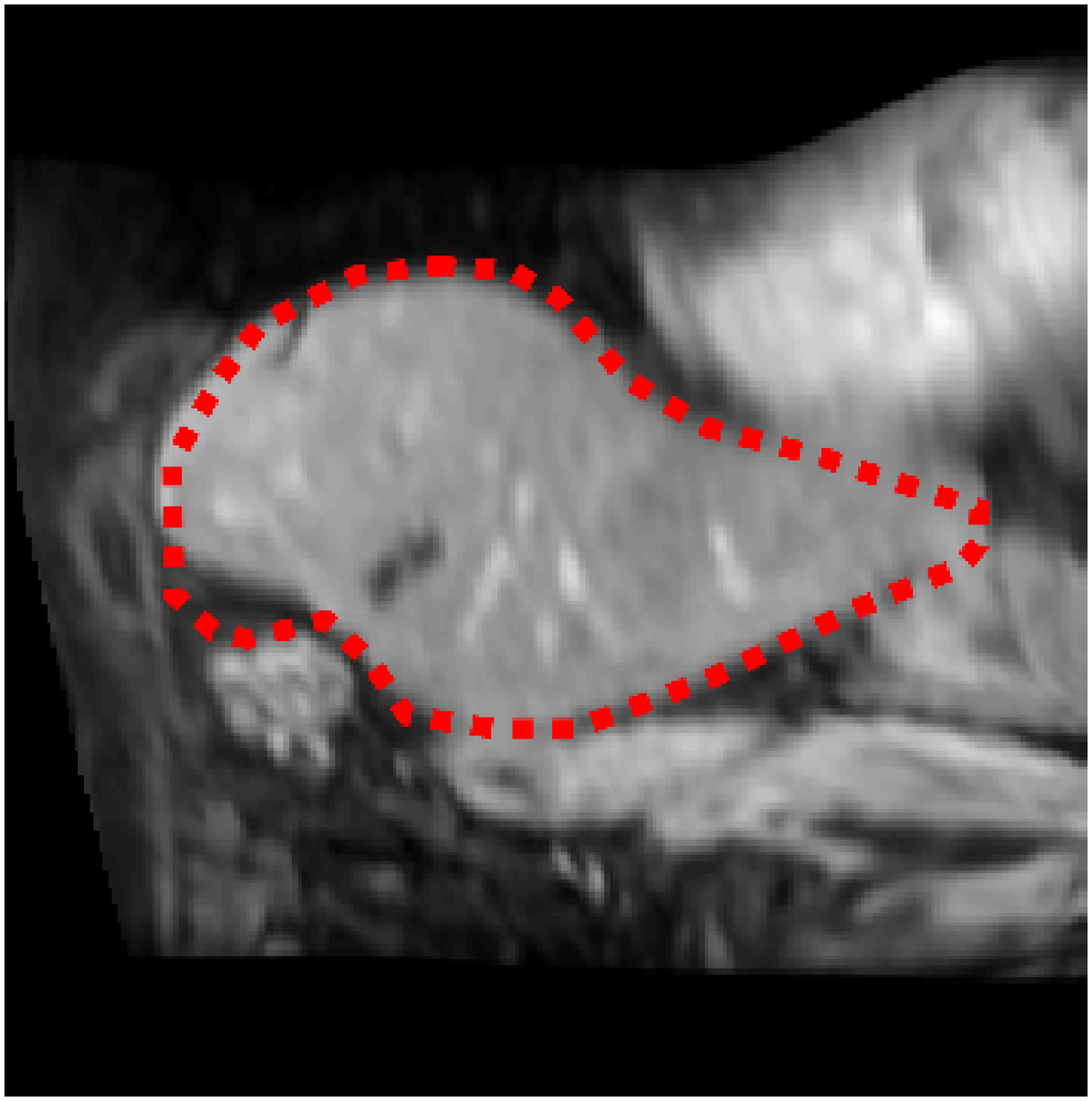}}
\centerline{(l)}\medskip
\end{minipage}

\begin{minipage}[b]{0.32\linewidth}
\centering
\centerline{\includegraphics[trim={0cm 0cm 0cm 0cm},clip,height=4.2cm]{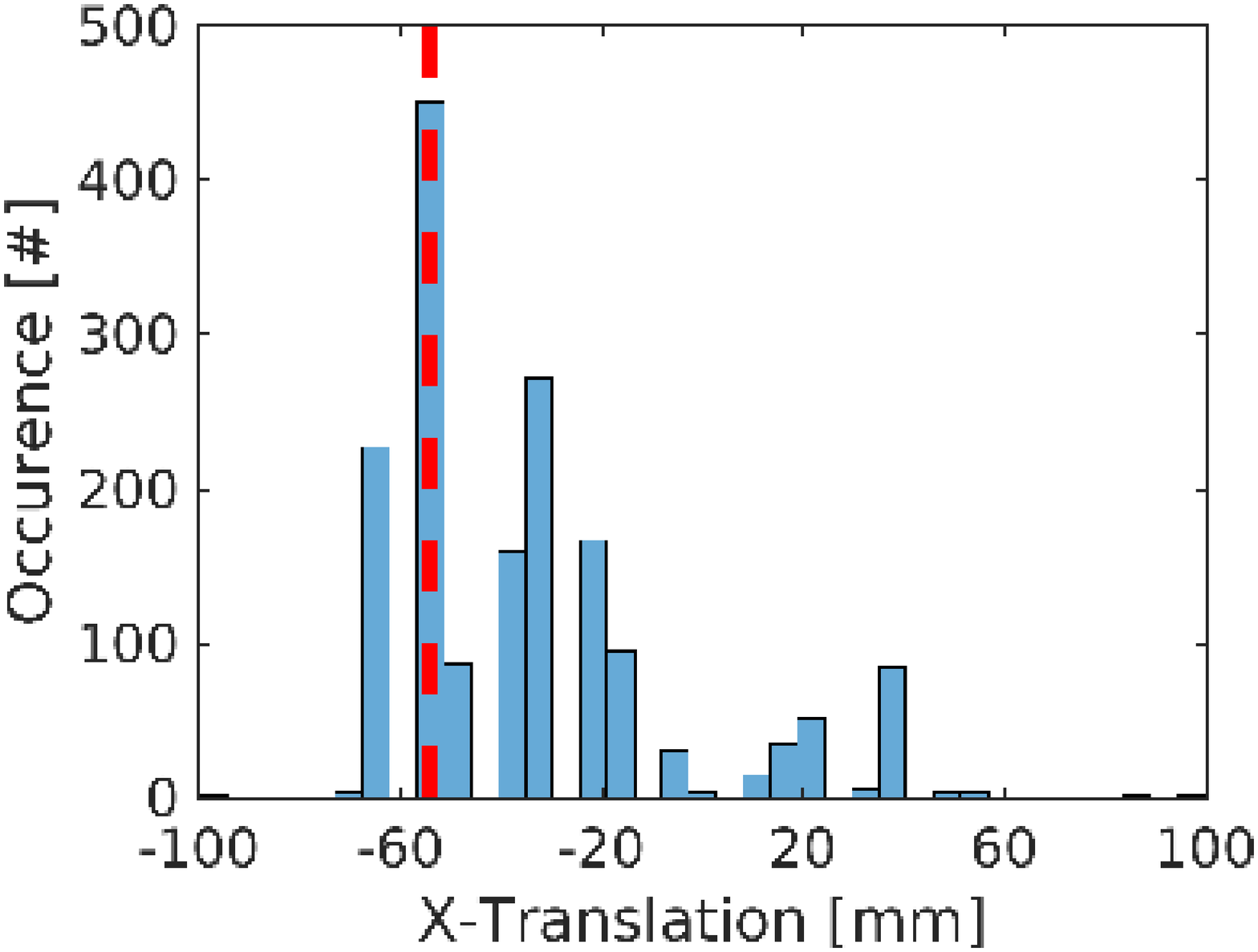}}
\centerline{(m)}\medskip
\end{minipage}
\begin{minipage}[b]{0.32\linewidth}
\centerline{\includegraphics[trim={0cm 0cm 0cm 0cm},clip,height=4.2cm]
{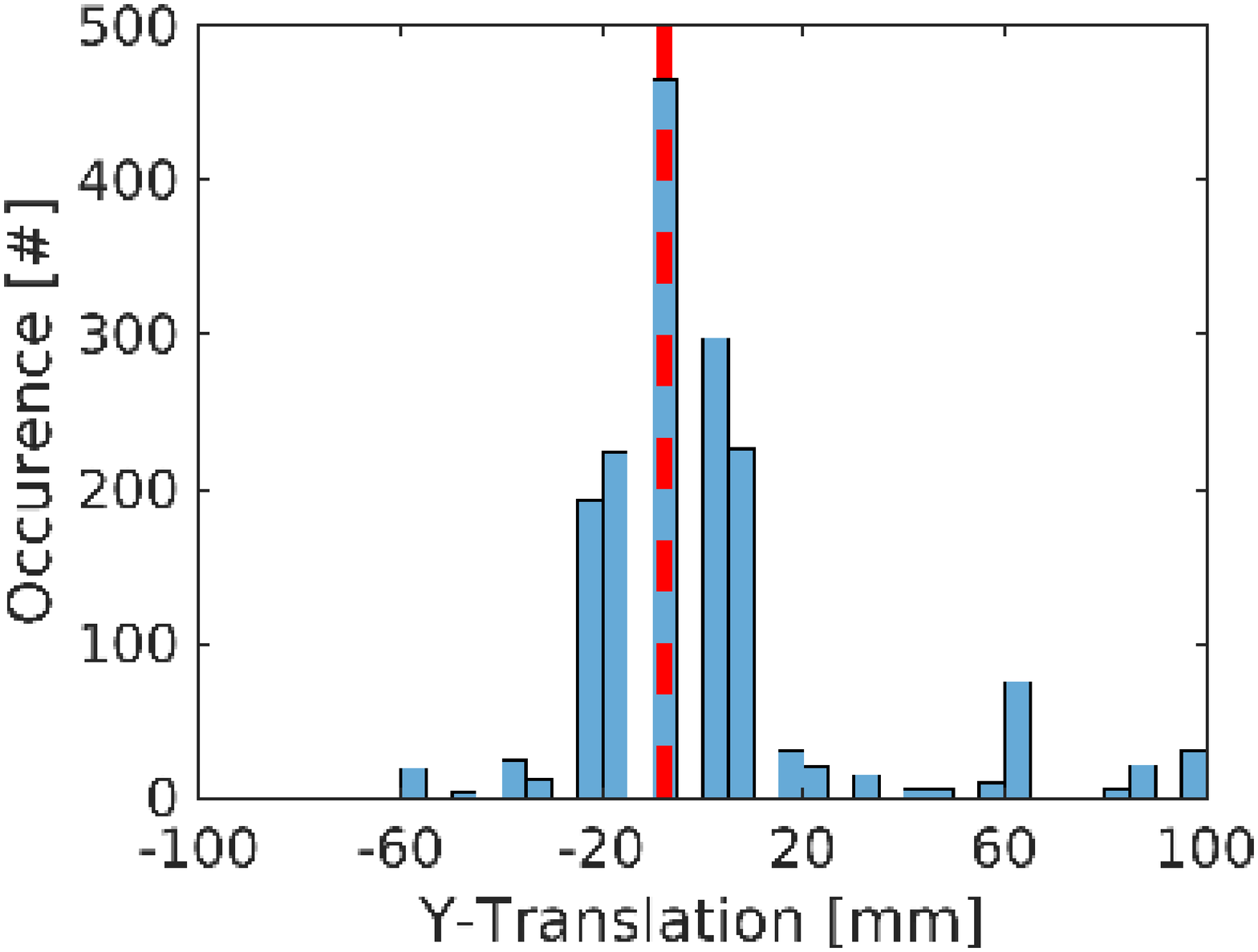}}
\centerline{(n)}\medskip
\end{minipage}
\begin{minipage}[b]{0.32\linewidth}
\centerline{\includegraphics[trim={0cm 0cm 0cm 0cm},clip,height=4.2cm]
{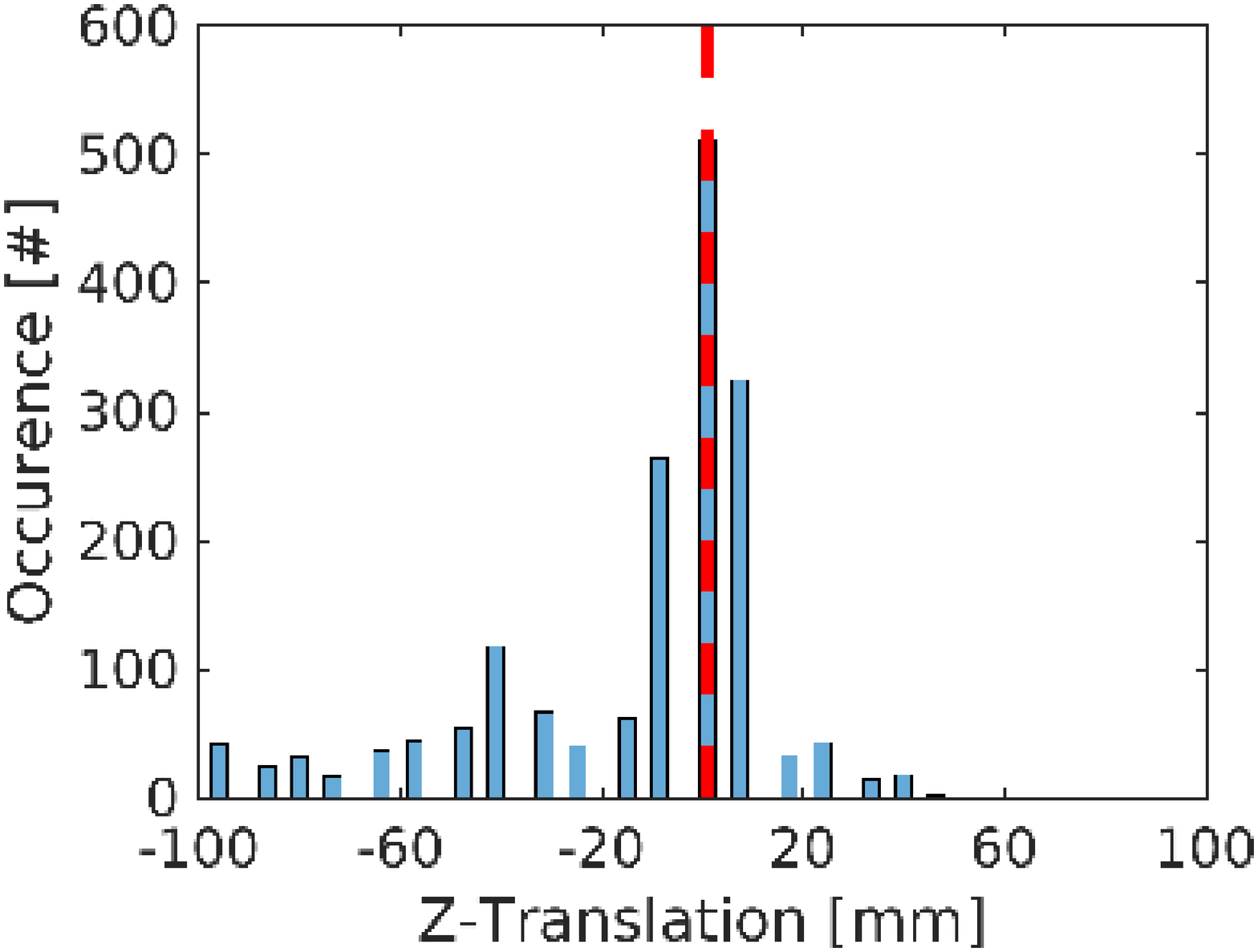}}
\centerline{(o)}\medskip
\end{minipage}

\caption{Example of a MR/CBCT registration results. The CBCT image, used as a reference for registration, was acquired immediately after insertion of 3 needles. Transversal (a-d), sagittal (e-h) and coronal (i-l) cross-sections are reported for: CBCT (first column), MRI before (second column) and after registration using PM-EA (third column) and PM-EA+Evo (fourth column). Histograms of X-, Y- and Z-shifts are reported in (m), (n) and (o), respectively (maximum occurrence in red dashed line).}
\label{fig:MRI_images}
\end{figure}

An example of CT/CBCT and MR/CBCT registration results are shown in figures \ref{fig:CT_images} and \ref{fig:MRI_images}, respectively. For both cases, the CBCT image (first column) was acquired intra-operatively after needle insertions and was employed as a reference for image registration. 
The pre-operative image is displayed before registration (second column), after PM-EA (third column) and after PM-EA+Evo (fourth column). The occurrence of patch shifts is reported for each spatial direction in panels (m--o): for each histogram, the shift with maximal occurrence is shown by the red dashed line. For panels (a--l), a ROI --- manually defined on the CBCT image/encompassing the liver --- is shown using red dash lines. Our visualization shows an improved correspondence of the contour of the liver with the manually defined liver boundary when the PM-EA solution is employed (see \ref{fig:CT_images}(c,g,k) and \ref{fig:MRI_images}(c,g,k)). Moreover, an even better correspondence of the contour is observable using the PM-EA+Evo solution (see \ref{fig:CT_images}(d,h,l) and \ref{fig:MRI_images}(d,h,l)).

\begin{figure}[h!]
\begin{minipage}[b]{0.49\linewidth}
\centering
\centerline{\footnotesize{CT/CBCT registration}}\medskip
\centerline{\includegraphics[trim={0cm 0cm 0cm 0.03cm},clip,height=5.6cm]{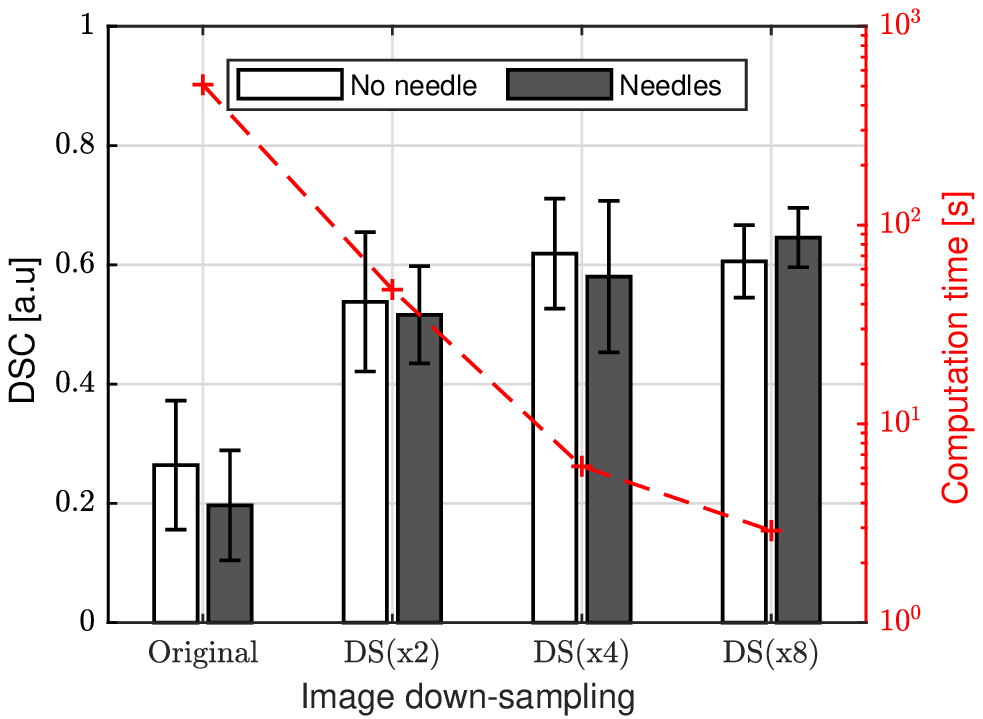}}
\centerline{(a)}\medskip
\end{minipage}
\begin{minipage}[b]{0.49\linewidth}
\centering
\centerline{\footnotesize{MR/CBCT registration}}\medskip
\centerline{\includegraphics[trim={0cm 0cm 0cm 0.03cm},clip,height=5.6cm]{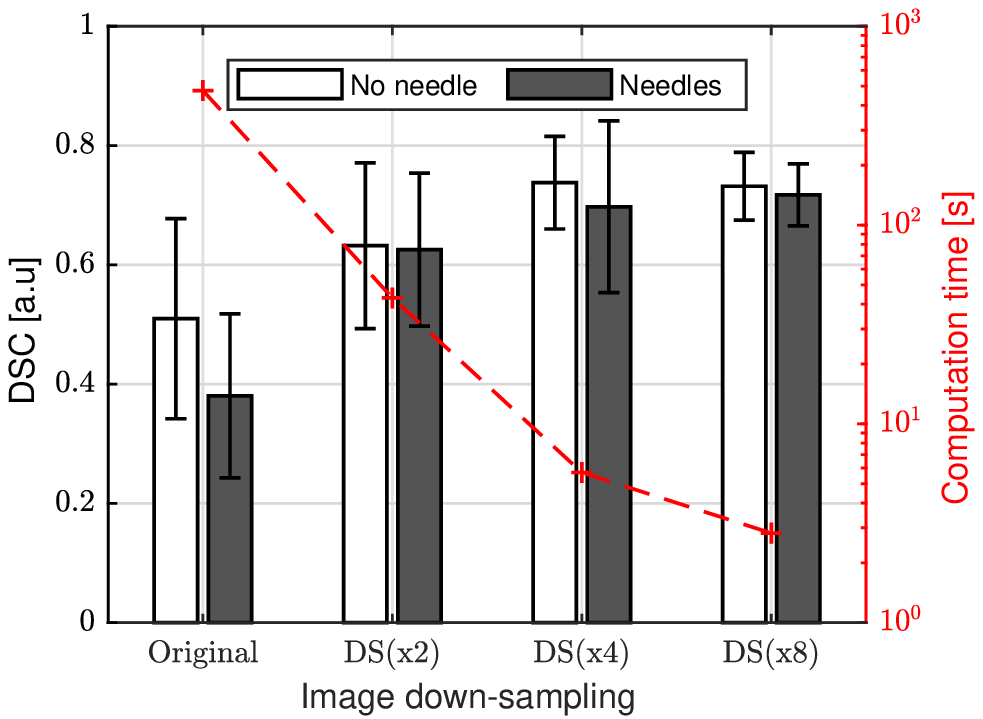}}
\centerline{(b)}\medskip
\end{minipage}
\caption{Analysis of the impact of the down-sampling of input data on the performance of the proposed PM-EA method. DSC (left Y-axis) and computation times (red dashed line/right Y-axis) are reported for the registration of CT/CBCT (a) and MR/CBCT (b) pairs for down-sampling factors $2\times$ (DS-2), $4\times$ (DS-4) and $8\times$ (DS-8). We recall that the patch size was here fixed to $9\times9\times9$ voxels. The number of histogram bins was fixed to a value of 50.}
\label{fig:DS_calibration}
\end{figure}

Figure \ref{fig:DS_calibration} analyzes the sensitivity to the down-sampling parameter (\emph{i.e.}, the input parameter $\#1$ of the proposed PM-EA method, see section \ref{sssec:calib}). A great DSC improvement together with a huge speed-up of the algorithm was obtained for increasing down-sampling factors. This tendancy was observed for both CT/CBCT (\ref{fig:DS_calibration}a) and MR/CBCT (\ref{fig:DS_calibration}b). Best results were obtained using the default down-sampling factor of $8\times$. For all tested scenarios, differences between DSC obtained before and after needle insertions were not statistically significant ($p\ge 0.2$ for CT/CBCT, $p\ge 0.3$ for MR/CBCT).

\begin{figure}[h!]
\begin{minipage}[b]{0.49\linewidth}
\centering
\centerline{\footnotesize{CT/CBCT registration}}\medskip
\vspace{0.15cm}
\centerline{\includegraphics[trim={0cm 0cm 0cm 0.03cm},clip,height=5.5cm]{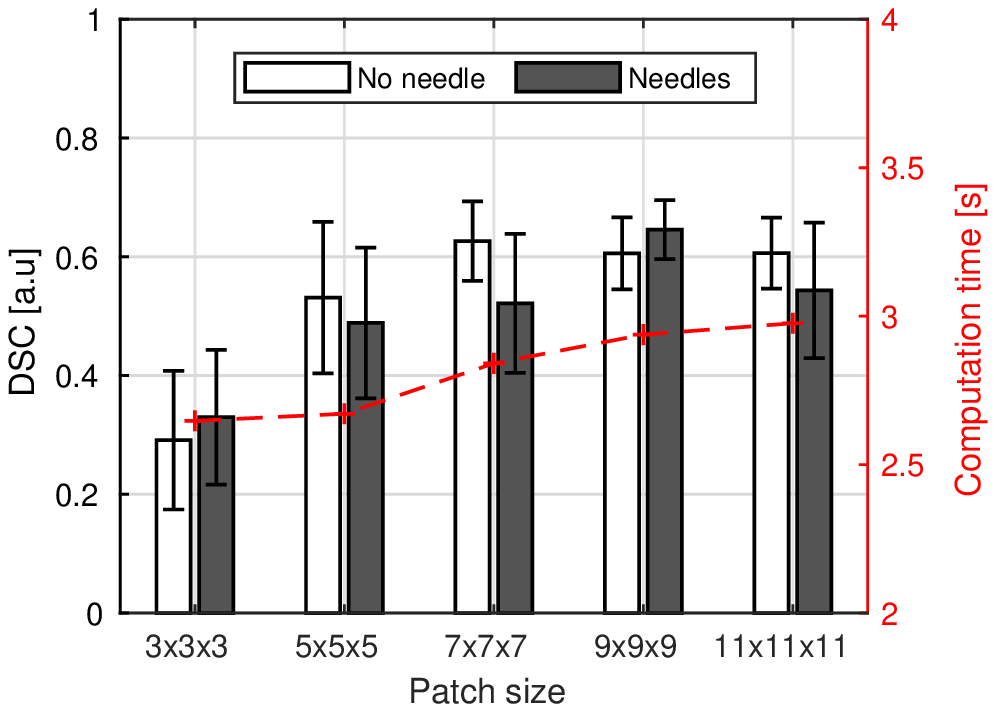}}
\centerline{(a)}\medskip
\end{minipage}
\begin{minipage}[b]{0.49\linewidth}
\centering
\centerline{\footnotesize{MR/CBCT registration}}\medskip
\vspace{0.15cm}
\centerline{\includegraphics[trim={0cm 0cm 0cm 0.03cm},clip,height=5.5cm]{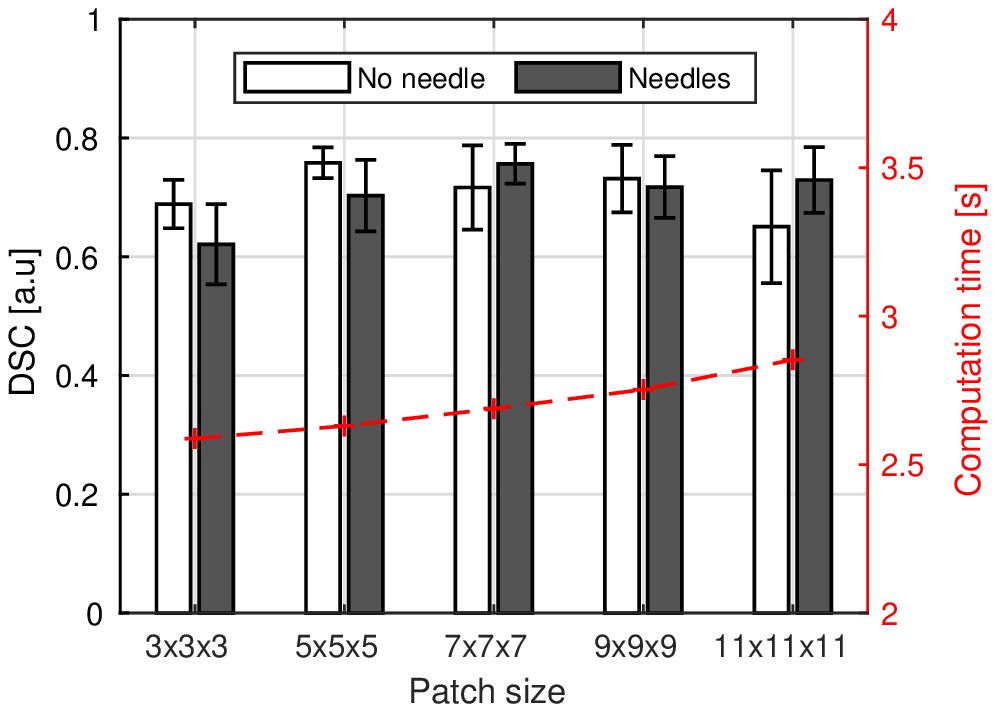}}
\centerline{(b)}\medskip
\end{minipage}
\caption{Analysis of the impact of the patch size on the performance of the proposed PM-EA method. DSC (left Y-axis) and computation times (red dashed line/right Y-axis) are reported for the registration of CT/CBCT (a) and MR/CBCT (b) pairs. We recall that the down-sampling factor of input data was here fixed to $4\times$. The number of histogram bins was fixed to a value of 50.}
\label{fig:PS_calibration}
\end{figure}

Figure \ref{fig:PS_calibration} analyzes the impact of the patch size (\emph{i.e.}, the input parameter $\#2$ of PM-EA). An improved registration accuracy with minimal losses in terms of computation time (several tenth of seconds) was obtained for an increasing patch size. This tendancy was observed for both CT/CBCT (\ref{fig:PS_calibration}a) and MR/CBCT (\ref{fig:PS_calibration}b). Best results were obtained using the default patch size of $9\times9\times9$ voxels. For all tested scenarios, differences between DSC obtained before and after needle insertions were not statistically significant ($p\ge 0.2$ for CT/CBCT, $p\ge 0.22$ for MR/CBCT).

\begin{figure}[h!]
\begin{minipage}[b]{0.49\linewidth}
\centering
\centerline{\footnotesize{CT/CBCT registration}}\medskip
\centerline{\includegraphics[trim={0cm 0cm 0cm 0.03cm},clip,height=5.6cm]{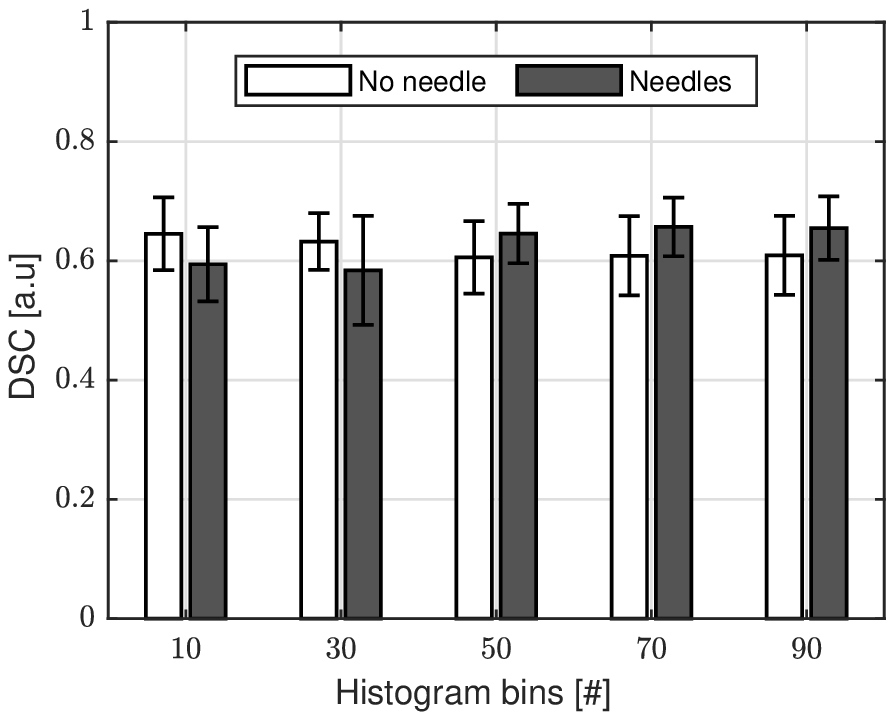}}
\centerline{(a)}\medskip
\end{minipage}
\begin{minipage}[b]{0.49\linewidth}
\centering
\centerline{\footnotesize{MR/CBCT registration}}\medskip
\centerline{\includegraphics[trim={0cm 0cm 0cm 0.03cm},clip,height=5.6cm]{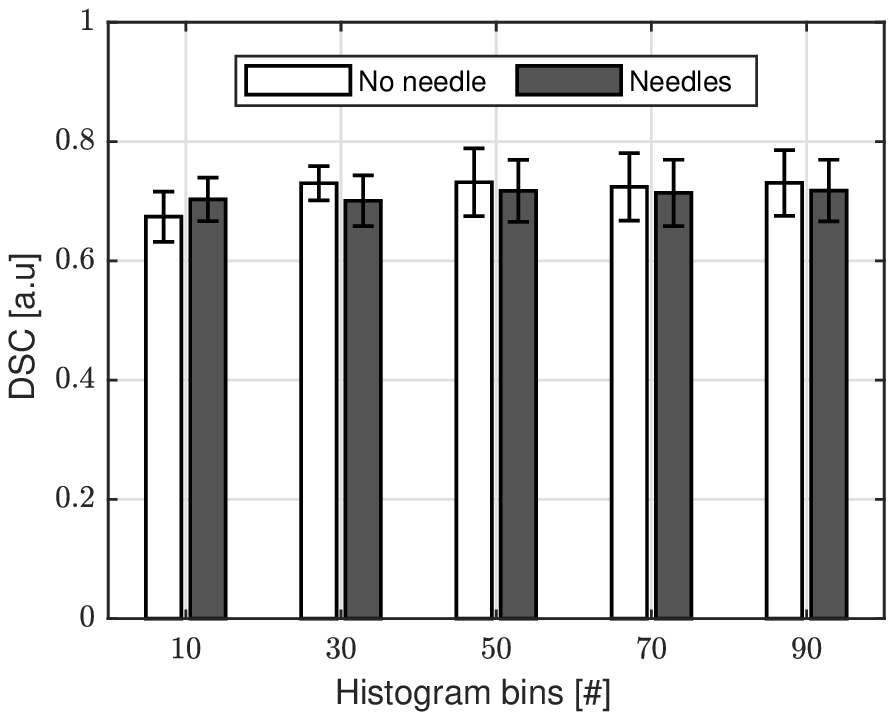}}
\centerline{(b)}\medskip
\end{minipage}
\caption{Analysis of the impact of the number of histogram bins on the performance of the proposed PM-EA method. DSC (left Y-axis) and computation times (red dashed line/right Y-axis) are reported for the registration of CT/CBCT (a) and MR/CBCT (b) pairs.  We recall that the down-sampling factor of input data was here fixed to $4\times$. The patch size was here fixed to $9\times9\times9$.}
\label{fig:HS_calibration}
\end{figure}

Differences between DSC for each pair of tested numbers of histogram bins (\emph{i.e.}, the input parameter $\#3$ of PM-EA) were not statistically significant ($p\ge 0.08$ for both CT/CBCT and MR/CBCT) (see figure \ref{fig:HS_calibration}). This was observable for  both CT/CBCT (\ref{fig:HS_calibration}a) and MR/CBCT (\ref{fig:HS_calibration}b). Here again, differences between DSC obtained before and after needle insertions were not statistically significant ($p\ge 0.11$ for CT/CBCT, $p\ge 0.16$ for MR/CBCT).

\begin{figure}[h!]
\begin{minipage}[b]{0.49\linewidth}
\centering
\centerline{\footnotesize{CT/CBCT registration}}\medskip
\vspace{0.15cm}
\centerline{\includegraphics[trim={0cm 0cm 0cm 0.03cm},clip,height=5.9cm]{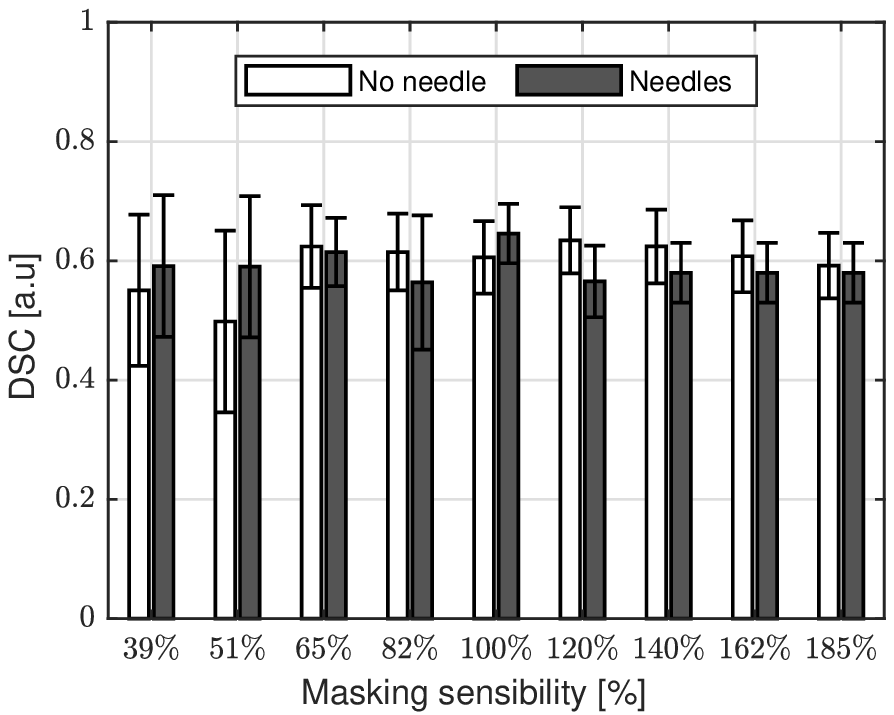}}
\centerline{(a)}\medskip
\end{minipage}
\begin{minipage}[b]{0.49\linewidth}
\centering
\centerline{\footnotesize{MR/CBCT registration}}\medskip
\vspace{0.15cm}
\centerline{\includegraphics[trim={0cm 0cm 0cm 0.03cm},clip,height=5.9cm]{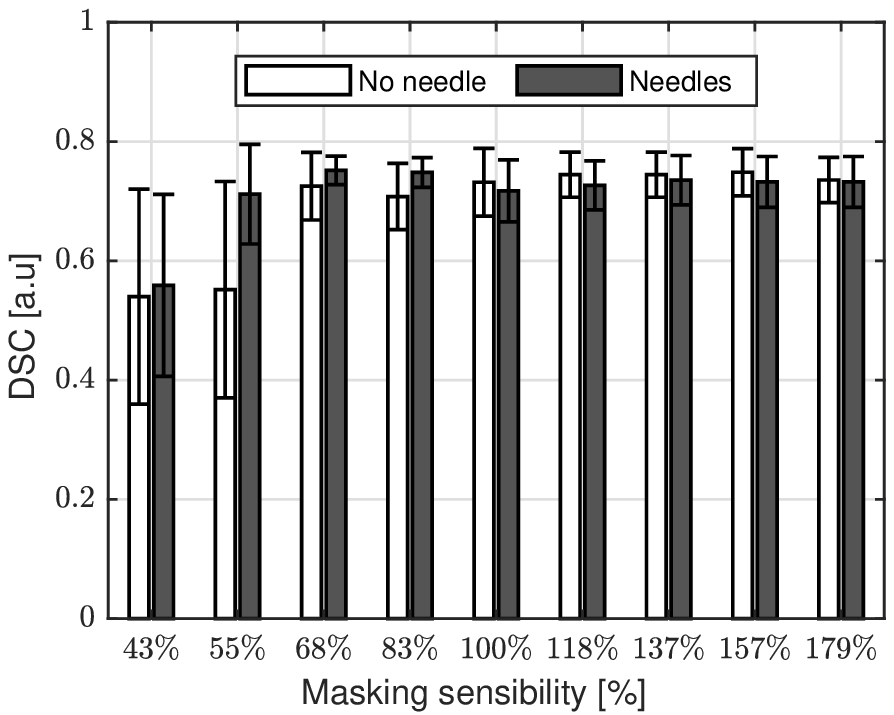}}
\centerline{(b)}\medskip
\end{minipage}
\caption{Analysis of the impact of errors occurred in the targeted organ delineation process performed on the pre-operative image $I$.}
\label{fig:Mask_calibration}
\end{figure}

Figure \ref{fig:Mask_calibration} analyzes the sensitivity of the proposed PM-EA algorithm against manual delineation errors of the organ of interest (\emph{i.e.}, the input parameter $\#4$ of PM-EA). A significant negative impact is observable for eroded versions of $M$, especially for MR/CBCT (\ref{fig:Mask_calibration}b) ($p\le 0.01$). Using dilated versions of $M$, DSC differences were not statistically significant ($p\ge 0.3$). Here again, differences between DSC obtained before and after needle insertions were not statistically significant ($p\ge 0.46$ for CT/CBCT, $p\ge 0.3$ for MR/CBCT).

\begin{figure}[h!]
\begin{minipage}[b]{0.49\linewidth}
\centering
\centerline{\footnotesize{CT/CBCT registration}}\medskip
\vspace{0.15cm}
\centerline{\includegraphics[trim={0cm 0cm 0cm 0.03cm},clip,height=5.8cm]{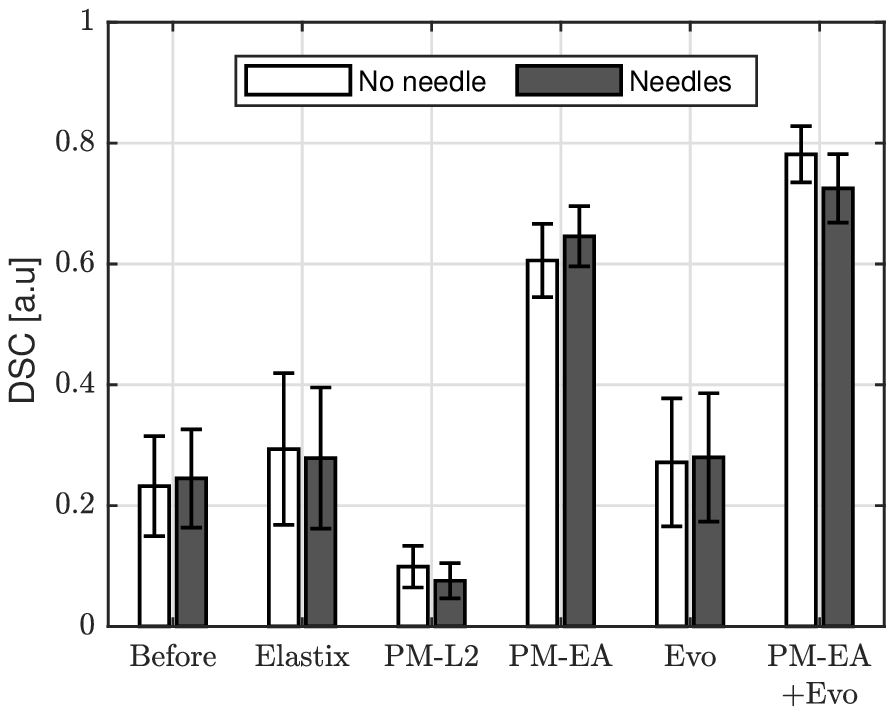}}
\centerline{(a)}\medskip
\end{minipage}
\begin{minipage}[b]{0.49\linewidth}
\centering
\centerline{\footnotesize{MR/CBCT registration}}\medskip
\vspace{0.15cm}
\centerline{\includegraphics[trim={0cm 0cm 0cm 0.03cm},clip,height=5.8cm]{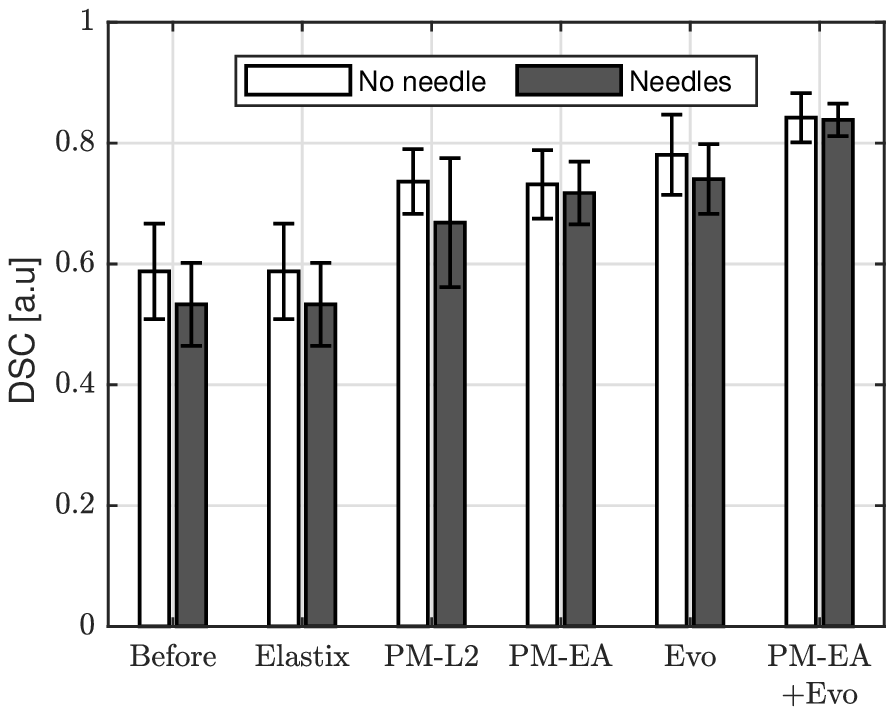}}
\centerline{(b)}\medskip
\end{minipage}
\caption{Summary of DSC scores obtained for the registration of CT/CBCT (a) and MR/CBCT (b) images, using tested solutions detailed in section \ref{sssec:tested_algos}. 
Standard deviations over the patients are given by the size of the black error bars. We recall that the down-sampling factor of input data was here fixed to $4\times$. The patch size was fixed to $9\times9\times9$ voxels. The number of histogram bins was fixed to a value of 50.}
\label{fig:DICEcomparison}
\end{figure}

Figure \ref{fig:DICEcomparison} compares the registration accuracy obtained using all tested algorithms (see section \ref{sssec:tested_algos} for details). 
Regarding solutions designed to estimate the global 3D shift of the liver (\emph{i.e.}, Elastix, PM-L2 and PM-EA), best results where achieved using the proposed PM-EA approach for both CT/CBCT (\ref{fig:DICEcomparison}a) and MR/CBCT (\ref{fig:DICEcomparison}b). PM-EA outperformed significantly a standard registration strategy implemented using the Elastix toolbox ($p=0.02$). The use of the proposed multi-modal metric improved significantly the DSC obtained using the L2-norm used in the original PatchMatch paper \cite{PatchMatch} ($p=0.01$). 
Regarding solutions designed to estimate the elastic deformation of the liver (\emph{i.e.}, Evo and PM-EA+Evo), the use of PM-EA improved significantly the performance of the tested multi-modal elastic registration algorithm Evo ($p=0.01$).
For all tested solutions, differences between DSC obtained before and after needle insertions were not statistically significant ($p\ge 0.2$ for CT/CBCT, $p\ge 0.08$ for MR/CBCT). It is interesting to note that the computational demand remained here below 30 seconds for the successive achievement of PM-EA and Evo algorithms with the used hardware for both CT/CBCT and MR/CBCT pairs.

\section{Discussion}

In the current study, we designed a novel method to estimate the global 3D translation between two multi-modal images. We retrospectively evaluate the proposed PM-EA algorithm under a realistic clinical scenarios. We focus on the specific interventional procedure of irreversible electroporation  (IRE) ablation for liver tumors. IRE technique provides an interesting alternative to standard ablative techniques, especially for tumor located near vital structures as detailed in \cite{Gallinato_2019}. Moreover, it gathers the main computational challenges in terms of medical image registrations, that have to be addressed to improve the procedures. Indeed, the procedures rely upon multimodal medical imaging: preoperative CT-scan or MRI to detect the target ablation region, and preoperative CBCT without and with needles to position the needles and to verify the positioning. Importantly, the needles positioning generate an elastic deformation of the liver, that has be accounted for as previously shown in \cite{Gallinato_2019}, and thus nonrigid multimodal algorithm as EVolution~\cite{EVolution2016} has to be used. As far as we know, the current non rigid registration algorithm needed an initial manual tuning  step to superimpose rouglhy the FOVs of two images of different modality. Importantly, the registration has be performed during the procedure, as demonstrated in  \cite{Gallinato_2019} in order to provide a numerical assessment of the therapy to the physicians. There is therefore a crucial need to automatize the image preprocessing of FOV alignement in any electroporation ablation procedures. In the current study, the use of various imaging sensors (CT/CBCT, MR/CBCT image pairs, CBCTs being acquired intra-operatively) is analysed as well as the impact of needle insertions during IRE procedures. Using the proposed experimental setup, the registration process is hampered by the use of different image FOVs, especially when using CT during the pre-operative session (see the low DCS obtained in figure \ref{fig:DICEcomparison} before registration when using CT instead of MRI). Moreover, partial FOVs, cross-contrast variations, appearing/disappearing (anatomical or not) structures are also involved between the image to register and the reference one.

Using such data sets, optimizing a simple translational model, as implemented in the Elastix toolbox, was found to be insufficient. The proposed regional registration approach using voxel patches provided a good structural compromise between the voxel-wise (as done with Evo) and ``global shifts'' (as done with Elastix in the scope of this study) approaches. Moreover, contrary to optimization methods which are inherently sensitive to local minima, PM-EA is able to deal with large translation amplitude, since potential patch matches are considered within the complete FOV, as described in section \ref{sssec::PM}. We have also shown that the proposed multi-modal image similarity metric, which favors edge alignements irrespective of the gradient direction, outperforms the L2-norm proposed in the original PatchMatch paper \cite{PatchMatch}. Ultimately, we have shown that PM-EA may greatly improve the performance of an existing multi-modal elastic registration algorithm (Evo in the scope of this study). 

As expected, computation times were greatly reduced using down-sampled versions of input data (see figure \ref{fig:DS_calibration}). This down-sampling step also acts as an inherent low-pass filter applied on $I$ and $J$ which improved the registration accuracy in our tests. Using the proposed default user-defined parameter (\emph{i.e.}, down-sampling factor of $8\times$), the average DSC with PM-EA exceeded 0.6 for both CT/CBCT, MR/CBCT image pairs together with a computation time cost below 3 seconds on a commodity hardware.

The proposed multi-modal metric of Eq. (\ref{eq:metric}) performs a weighted average over patches of the edge alignement score. Consequently, the patch size is an input parameter which can be increased in order to mitigate the fact that the observed image features might not be discriminative enough. Increasing the patch size may thus improve the robustness against anatomical structure without counterpart between the image to register and the reference one. This benefit was achievable with a moderate negative impact on the computation time. However, to some extent, increasing the patch size may be unable to cope with complex local tissue deformations. A good compromise in the choice of the patch size is thus essential for a reliable and accurate patch matching. Using the proposed default user-defined parameter (\emph{i.e.}, patch size of $9\times 9\times 9$ voxels), PM-EA attained the best results in terms of DSC for all tested experimental conditions (CT/CBCT, MR/CBCT registration, needle insertions), as shown in figure \ref{fig:PS_calibration}.

It can be noticed that the number of histogram bins had no impact on the overall results, as shown in figure \ref{fig:HS_calibration}. The default user-defined parameter (\emph{i.e.}, 50 bins) was thus well suited for all presented results.

In our data, the liver undergoes complex deformations between the images being registered. The registration accuracy decreased for eroded versions of $M$, as shown in figure \ref{fig:Mask_calibration}. Liver boundaries, which are needful contrast regions, are not taken into account in such a case. Moreover, the overall shift estimate is likely to differ from the global liver displacement if the manually defined mask $M$ only includes a subset of the liver. Alternatively, no significant impact on the registration accuracy was observed for dilated versions of $M$ in our tests. Therefore, the guideline is that $M$ must include at least the targeted organ.

\section{Conclusion}

The successfull completion of an interventional therapeutic workflow often relies on establishing a spatial coherence between images acquired by various sensors at different stages. 
The proposed PM-EA algorithm was validated in several complementary experiments. It was demonstrated that it outperforms existing registration solutions for the estimation of a global 3D translation between two multi-modal images. 
The method can be used as a pre-conditioning step for a more complex multi-modal elastic registration algorithm.
The method was thereby beneficial for CT to CBCT and MRI to CBCT registration tasks, especially when highly different image FOVs are involved. In addition, this was achievable together with a computation time cost below 3 seconds on a commodity hardware using our experimental protocol. The proposed patch-based workflow thus represents an attractive asset for DIR at different stages of an interventional procedure.

\section*{Acknowledgment}

Experiments presented in this paper were carried out using the PlaFRIM experimental testbed, supported by Inria, CNRS (LABRI and IMB), Universit\'e de Bordeaux, Bordeaux INP and Conseil R\'egional d'Aquitaine (see https://www.plafrim.fr/). The authors thank the Laboratory of Excellence TRAIL ANR-10-LABX-57 for funding. This study has been carried out with the financial support of the French National Research Agency (ANR) in the frame of the ``Investments for the future'' Programme IdEx Bordeaux-CPU (ANR-10-IDEX-03-02). This research has been partly granted by the Plan Cancer project NUMEP (Inserm 11099), led by C.P.

\section*{References}

\bibliographystyle{dcu}
\bibliography{2020_Patch_Match_registration_final}

\end{document}